\documentclass[aps,onecolumn,nofootinbib,groupedaddress,amsfonts,floatfix]{revtex4-1} 

\usepackage{tikz}
\usepackage{hyperref}
\usepackage{amsmath}
\usepackage{tensor}
\usepackage{graphicx}
\usepackage{textcomp}
\usepackage{amsfonts}
\usepackage{bm}
\usepackage{subcaption}
\usepackage{tabularx}
\usepackage{pgfplots}
\usepackage{mathrsfs}
\usepackage{mathtools}
\usepackage{soul}
\usepackage{float}

\graphicspath{{Figures/}}

\usepackage{color}

\newcommand{\dotsize}{2pt}
\newcommand{\arcwidth}{2}
\newcommand{\gapwidth}{1}
\newcommand{\nn}{\nonumber \\}
\newcommand{\eqn}[1]{Eqn. \ref{#1}}

\newcommand{\Square}[1]{+(-#1,-#1) rectangle +(#1,#1);}
\newcommand{\NodesUp}[3]{
 \fill (#2,#3) circle[radius=\dotsize];
 \draw (#2,#3) node[below]{0};
 \ifnum #1 > 0
  \foreach \x in {1, ..., #1} {
   \fill (\arcwidth*\x/#1+#2,#3) \Square{\dotsize};
   \draw (\arcwidth*\x/#1+#2,#3) node[below]{\x};
  };
 \fi 
}
\newcommand{\NodesDn}[3]{
 \fill (#2,#3) circle[radius=\dotsize];
 \draw (#2,#3) node[above]{0'};
 \ifnum #1 > 0
  \foreach \x in {1, ..., #1} {
   \fill (\arcwidth*\x/#1+#2,#3) \Square{\dotsize};
   \draw (\arcwidth*\x/#1+#2,#3) node[above]{\x'};
  };
 \fi 
}
\newcommand{\DrawLine}[6]{
 \ifnum #5 > 0
  \def\a{(\arcwidth*#5/#1+#3,#4)};
 \else
  \def\a{(-\arcwidth*#5/#2+#3,-\gapwidth+#4)};
 \fi
 \ifnum #6 > 0
  \def\b{(\arcwidth*#6/#1+#3,#4)};
 \else
  \def\b{(-\arcwidth*#6/#2+#3,-\gapwidth+#4)};
 \fi
 \draw [black,dotted] \a -- \b;
}

\newcommand{\CouplingUpR}[4]{
 \NodesUp{#1}{#2}{#3};
 \foreach[count=\i] \couple in {#4} {
  \def\b{#1}
  \ifx\i\b
   \draw [black] (\arcwidth*\i/#1+#2,#3) arc (0:180:\arcwidth/#1/2*\couple);
  \else
   \draw [black] (\arcwidth*\i/#1+#2,#3) arc (0:180:\arcwidth/#1/2*\couple);
  \fi
 };
}

\newcommand{\CouplingUp}[4]{
 \NodesUp{#1}{#2}{#3};
 \ifnum #1 > 0
  \foreach[count=\i] \couple in {#4} {
   \draw (\arcwidth*\i/#1+#2,#3) arc (0:180:\arcwidth/#1/2*\couple);
  };
 \fi
}
\newcommand{\CouplingDn}[4]{
 \NodesDn{#1}{#2}{#3};
 \ifnum #1 > 0
  \foreach[count=\i] \couple in {#4} {
   \draw (\arcwidth*\i/#1+#2,#3) arc (0:-180:\arcwidth/#1/2*\couple);
  };
 \fi
}

\newcommand{\nh}{\hat{{\bf n}}}
\newcommand{\na}{\hat{\alpha}}
\newcommand{\tn}{\Theta(\hat{{\bf n}})}
\newcommand{\dd}{\mathrm{d}}
\newcommand{\bx}{{\bf x}}
\newcommand{\bk}{{\bf k}}
\newcommand{\opa}[1]{\nabla_{\hat{{\bf n}}}^{#1}}
\newcommand{\opb}[2]{(\hat{\Box}_{\nh,#1})_{#2}}
\newcommand{\opc}[1]{\nabla_{#1}^{\nh}}
\newcommand{\opd}[1]{D_{#1}}
\newcommand{\NC}[1]{[#1]}

\begin{document}
{\hfill KCL-PH-TH/2021-67}

\title{Weak lensing ``post-Born'' effects are equivalent to pure lens-lens couplings}

\author{Oliver Denton-Turner${^\dagger}$ and Eugene A. Lim${^\dagger}$}
\email{eugene.a.lim@gmail.com}
\affiliation{${^\dagger}$Theoretical Particle Physics and Cosmology Group, Physics Department,
Kings College London, Strand, London WC2R 2LS, United Kingdom\\}
\begin{abstract}
We show that the so-called ``post-Born'' effects of weak lensing at 4th order are equivalent to lens-lens couplings in the Born Approximation. We demonstrate this by explicitly showing the equivalence of the canonical weak lensing approach at 4th order using the anisotropy remapping method, to that of the 4th order calculation of the lens-lens coupling effects using the Boltzmann equation approach that was first developed in \cite{Su:2014mga}. Furthermore, we argue that to incorporate true ``post-Born'' effects, i.e. taking into account non-straight photon paths, require the addition of a photon deflection term which has not been taken into account in the canonical formalism nor the Boltzmann method.

\end{abstract}
\maketitle

\section{Introduction}
Weak lensing of the cosmic microwave background (CMB) is now well established as an important probe of cosmological parameters \cite{Pyne:1995bs,Kaiser:1996tp,Wu:2019hek,Namikawa:2019gtc,Aghanim:2018oex,1987A&A...184....1B,Zaldarriaga_1997,Lewis_2005,Kaplinghat_2003,Liu:2016nfs,Green:2016cjr}. By using the observations of the temperature $\Theta$, and the $E$ and $B$ polarisation anisotropies spectra of the CMB, we can reconstruct the underlying lensing potential \cite{Bartelmann:1999yn,Hanson_2011,Lewis_2011,Zaldarriaga:1998te,Hu:2001tn,Hanson_2010,Peloton:2016kbw,Carron:2017mqf} $\phi(\nh) =-2\int_0^{r_s} dr_1 W(r_1,r_s)\Psi_W$ where $W(r_1,r_2) \equiv 1/r_1-1/r_2$ is the lensing efficiency kernel \cite{Bartelmann:1999yn,Cooray:2002mj},  $\Psi_{W}$ is the Weyl potential and $r_s$ is the conformal distance to the last scattering surface (LSS). The lensing potential contains the projected information of the foreground distribution of matter, and hence the lensing potential allows us to probe its correlations \cite{Aghanim:2018oex,Song_2003,Hand_2015,Singh_2016,Harnois_D_raps_2017,Kirk_2016,Liu_2015,Seljak_1999,Goldberg_1999,Namikawa:2016jff,Nguyen:2017zqu,Polarbear:2019zli,Das:2013aia,Larsen:2015aoa,2017PhRvD..95l3512S,2012ApJ...759...32V}.

To do so, a formalism to map the CMB anisotropies to $\phi(\nh)$ and implicitly $\Psi_W(r)$ must be constructed. The standard canonical method focuses on the computation of the deflection angle $\alpha^a(\nh)=\partial^a\phi(\nh)$ which can then be directly mapped order by order onto the observed $\Theta$ spectra \cite{Okamoto:2003zw} using the following \emph{ansatz}, i.e.
\begin{eqnarray}
\Theta(\nh + \na)& \approx& \Theta(\nh) + \alpha^a\opc{a}\tn + \frac{1}{2}\alpha^a\alpha^b\opc{a}\opc{b}\tn + \dots \nn
&\equiv&\Theta^{[I]}(\nh) + \Theta^{[II]}(\nh)+\Theta^{[III]}(\nh)+\dots.\label{eqn:theta_expansion_intro}
\end{eqnarray}
This remapping can be carried out to increasingly higher orders in $\Theta$ \cite{Krause:2009yr,Cooray:2002mj,Pratten:2016dsm,Fanizza:2015swa,Marozzi:2016uob}. 
While these effects are expected to be small, they may play a role in the estimates of systematics of higher order correlations \cite{Fabbian:2019tik} thus an accurate accounting of them is required. 

In such calculations, the Born Approximation is applied -- the line integral from the observer to the CMB source is assumed to be through the background unperturbed photon path \cite{Lewis:2006fu}.  In the calculation of the 4th order lensing anisotropy, it is often suggested that the approximation no longer holds, and hence the physical effects are attributed to ``post-Born'' physics \cite{Pratten:2016dsm,Marozzi:2016uob,Marozzi:2016und} such as ray-deflection and time delay. In this paper, we argue that in such calculations, the Born Approximation still holds -- in other words the line integrals are still taken along the background unperturbed photon path\footnote{A numerical approach is undertaken by  \cite{Fabbian:2017wfp} which does take into account the ray deflection effects.}. Furthermore, we will show that these effects commonly attributed as ``post-Born'' are generated by higher order lens-lens couplings instead.

To show this, we will use an alternative method for calculating the effect of weak lensing on the CMB temperature anisotropies first introduced in \cite{Su:2014mga} which is to directly solve the Boltzmann equation for the lensed anisotropies order by order. This is in contrast to solving the geodesic equation to obtain the deflection angle as an intermediary step, as in the case of the canonical method, and then mapping those onto the anistropies via \eqn{eqn:theta_expansion_intro}.  In \cite{Su:2014mga}, the lensing anisotropies at 4th order due to pure lens-lens couplings were calculated, and in this paper, we will show that the results obtained in that paper is completely equivalent to that of the canonical method, and hence proving our assertion. Furthemore, we will identify the point of departure where true post-Born effects such as ray deflection should be introduced.

The paper is organized as follows. In Section \ref{sect:boltzmann} we will describe the Boltzmann formalism as introduced in \cite{Su:2014mga}, and identify the point where true post-Born effects could be introduced. In Section \ref{equivalence}, we will clarify the relationship between the the Boltzmann approach and the canonical approach, and explicitly prove the equivalence at 4th order, i.e. the results obtained in \cite{Pratten:2016dsm, Marozzi:2016uob} are completely equivalent to that first obtained in \cite{Su:2014mga}. Finally in Section \ref{sect:conclusion} we conclude.

\section{The Boltzmann  Equation Approach} \label{sect:boltzmann}

We now forego any \emph{ansatz} and obtain all the higher order $\Theta$ terms without any reference to the deflection angles by explicitly solving its Boltzmann evolution equation \cite{Pitrou:2008hy}. This approach was introduced in \cite{Su:2014mga} and used to calculate effects of lens-lens couplings on weak lensing to 4th order, and which this section closely follows. 

The Boltzmann equation describes the evolution of a distribution of photons, i.e. the intensitiy $I$, as a function of spactime position $x^{\mu}$, frequency $p^0$ and spatial direction $\nh$
\begin{equation}
{\cal L}[I(x^{\mu},p^0,\nh)]= 2{\cal C}(x^{\mu},p^0,\nh)~,
\end{equation}
where ${\cal L}$ is the Liouville operator which defines its free evolution ${\cal L} = \mathrm{d}/\mathrm{d}\eta$, with $\eta$ being the conformal time.  It is a first order \emph{total derivative}, encapsulating much of the physics of the evolution despite its notational simplicity.  Meanwhile, ${\cal C} = - 1/2\tau' I+{\cal D}$ is the collision term where $\tau' = \partial_{\eta} \tau$ is the time derivative of the optical depth and ${\cal D}$ is the Compton scattering term.

The temperature anisotropy $\Theta$ is directly related to the energy averaged intensity $\hat{I}$ and hence $I$ by the following relation
\begin{equation}
\Theta = \frac{1}{4}\hat{I}~,~ \hat{I}(x^{\mu},\nh) = \frac{\int I(x^{\mu},p^0,\nh)(p^0)^3dp^0}{\int {\cal I}(p^0)(p^0)^3 dp^0}~, \label{eqn:theta_I_relation}
\end{equation}
where ${\cal I}$ is the isotropic background blackbody spectrum.  The Boltzmann equation serves as the evolution equation for $\Theta$, thus we can expand $\hat{I}$ as usual $\hat{I} = \sum_N \hat{I}^{[N]}/N!$ which is related order by order to the expansion $\Theta = \sum_N \Theta^{[N]}$ by $\Theta^{[N]} = \hat{I}^{[N]}/4N!$.

Expanding the Liouville term ${\cal L}$,
\begin{eqnarray}\label{eqn:Liouville_intensity}
{\cal L} = \left[ \frac{\partial}{\partial \eta} + {\mathrm{d}x^{a} \over \mathrm{d} \eta} \opd{a} + {\mathrm{d}p^0 \over \mathrm{d} \eta} {\partial \over \partial p^0}
+ {\mathrm{d}n^a \over \mathrm{d} \eta}\opc{a}\right],
\end{eqnarray}
where $\opd{a}$ is the spatial covariant derivative along the photon trajectory $x^a$ and $p^a \equiv \dd x^a/\dd\eta$ is its first derivative. The first term is the overall time evolution of the intensity $\partial_{\eta} I$ and the third term denotes the redshifting of the photon energy -- for now we will only keep this term at 1st order in perturbation (as shown in \cite{Su:2014mga}, higher order terms are subdominant).  

The fourth term encodes the weak lensing effect -- at first order, it is given by the familiar spatial geodesic equation $(\mathrm{d}p^a / \mathrm{d} \eta)^{[I]} = -2\delta^{ab}\partial_{b}\Psi_{W}$.  However, since the intensity is a function of $p^a$, \emph{we do not need to directly integrate this equation} -- the physics of geodesic deviation is automatically folded into the formalism.  Indeed since we assume that there is no backreaction of the perturbed photon intensity to the metric\footnote{To incorporate this effect, we would need to solve the perturbed Einstein equation to obtain the 2nd order metric perturbations. This effect is also ignored in the standard canonical formalism.}, the fourth term of  \eqn{eqn:Liouville_intensity} \emph{to all orders} is \cite{Su:2014mga}
\begin{equation}
\frac{\mathrm{d}n^a}{\mathrm{d}\eta} = -2S^{ab}\partial_b\Psi_W~,
\end{equation}
where $S_{\mu\nu} = g_{\mu\nu} - n_{\mu}n_{\nu}$ projects  onto the hypersurfaces labeled by the cosmic time $\eta$ such that the co-moving observer measures the photons at energy $p^0$. We will assume that the spatial components $S^{ab}\rightarrow \delta^{ab}$ to match to the fact that in the remapping ansatz, $\nh + \hat{\bf \alpha}$ is assumed to be a unit vector (even though in principle it is not).

Finally, the second term  $\dd x^a/\dd \eta$ captures deviation of the photon trajectory from the unperturbed background path $\bar{\bf x}$. A hint can be gleaned by considering the first order term of its coefficient
\begin{equation}
\left(\frac{d x^a}{d\eta}\right)^{[I]}=2n^a\Psi_W~, \label{eqn:1stordertimedelay}
\end{equation}
i.e. it depends on $\Psi_W$ and \emph{not its spatial derivative}, unlike the lensing term.  This fact is true to all orders in perturbation theory.  If we ignore all the other interactions of the Liouville operator and focus on this term and the evolution term, at first order it is
\begin{equation}
\frac{\partial}{\partial \eta}I = \frac{dx^a}{d\eta}D_a I~,
\end{equation}
which can be written as
\begin{equation} \label{eqn:time_delay_1st}
\left(\frac{\partial}{\partial \eta}-2\Psi_W n^aD_a\right)I =0~,
\end{equation}
describing an integral along the deflected photon path.  Simply put, \eqn{eqn:1stordertimedelay} is an equation of motion for $\bx$ as a function of $\Psi({\bx})$ thus a full accounting of it will encode the effects of deviating from the background path $\bar{\bx}$.  An equivalent interpretation of \eqn{eqn:1stordertimedelay} is that since the RHS depends purely on $\Psi_W$ and not its spatial derivative, it captures the \emph{time delay} of the photon path due to the presence of perturbations \cite{Fanizza:2018qux}. 

In principle, this term should be considered -- it represents ``post-Born'' effects in the strictest sense as including this term would mean that $\Psi_W$ is now a function of $\bx$ (and not just $\bar{\bx}$), capturing deviations from the unperturbed path.  This term is \emph{not} captured by the canonical formalism since \emph{all} quantities in this formalism is evaluated at the background path $\bar{\bx}$ -- note that the photon path perturbation $\delta \bx$ itself is derived by solving the geodesic equation order by order using the Weyl potential evaluated at $\bar{\bx}$. As we will see below, by dropping this term, we will recover an equivalent relationship between the Boltzmann approach and the remapping ansatz. 

Given these approximations, and integrating over the energy \eqn{eqn:theta_I_relation}, the Boltzmann equation for $\Theta$ to $N$-th order is\footnote{For a detailed derivation, see \cite{Su:2014mga}.} 
\begin{equation}
\left[\partial_{\eta} + n^a \partial_a
 +\tau^{\prime} \right] \Theta^{[N]}
 = \hat{{\cal D}}^{[N]}-2\opa{a}\Psi_W\opc{a}\Theta^{[N-1]} ~.
\label{eqn:BE_1}
\end{equation}
where $\opa{a} \equiv \delta^{ab}\opc{b}$. Focusing on lensing, we will ignore Compton scattering terms beyond first order, i.e. $\hat{{\cal D}}^{[N]}=0$ for $N>1$. This hierarchy of equations can be solved iteratively using standard perturbation theory as follows.

Transitioning to Fourier space, the first order solution is obtained from the familiar line-of-sight integral \cite{Seljak:1996is}
\begin{eqnarray}
\Theta^{[I]}(\eta, {\bf{k}}, \hat{\bf{n}}) =
e^{\tau(\eta)}\int^{\eta}_{0} \mathrm{d}\tilde{\eta} ~S_T e^{i{\bk\cdot\nh}(\tilde{\eta} - \eta)}~, \label{eqn:lensing_source}
\end{eqnarray}
where the source function $S_T(\tilde{\eta},\Psi_W, \bk\cdot\nh/k)$ describes the usual 1st order collision term physics \cite{Dodelson:2003ft}, and $\eta=0$ is the time of mode reentry.  

At $N>1$ order, the Boltzmann equation \eqn{eqn:BE_1}  can be directly integrated in Fourier space using the standard line-of-sight integral to obtain the rather intimidating recursive equation
\begin{widetext}
\begin{align}
e^{-i\bk_{N+1}\cdot\nh r_{N+1}-\bar{\tau}(\eta_{N+1})} \Theta^{[N+1]}(\eta_{N+1},\bk_{N+1},\nh) = 
\int_0^{\eta_{N+1}} d\eta_{N}\int\frac{d\bk_N'd\bk_N}{(2\pi)^{3/2}} \delta(\bk_{N+1}-\bk_{N}-\bk_{N}')\\
\times \opa{a}\left(e^{-i\bk'_N\cdot\nh r_N}\Psi_W(\bk_N')\right)\underbrace{\left[-\left(\frac{2}{r_N}\right)\left(\opc{a}+ik_{N,a}r_N\right)\right]}_{\opb{r_N}{a}}\left(e^{-i\bk_{N}\cdot\nh r_N-\bar{\tau}(\eta_N)}\Theta^{[N]}(\bk_N,\eta_N,\nh)\right)~, \label{eqn:BE_4}
\end{align}
\end{widetext}
where $\bar{\tau}(\eta_N) \equiv \int_{\eta_N}^{\eta_0}\tau'(\bar{\eta})d\bar{\eta}$ and 
$r_N\equiv \eta_0-\eta_N$, with $\eta_0$ being the conformal time today, and $\bar{\bf x} = \nh r$. To make contact with the notation in the remapping approach,  we define $r_s = \eta_0-\eta_{LSS}$, where $\eta_{LSS}$ is the time of recombination. 

The key operator in the above equation is the underbraced \emph{lensing operator} $\opb{r}{a}$, which generates a lensing kernel $W(r,\tilde{r})$ when acting on a source or lensing potential, 
\begin{equation}
\int \frac{d^3\bk}{(2\pi)^{3/2}}\opb{\tilde{r}}{a}\left(e^{-i\bk\cdot\nh r} \Psi_W(\bk)\right)=
-2W(r,\tilde{r})r\partial_a\Psi_W({\bar{\bx}})~.
\label{eqn:lensing_op}
\end{equation}
\eqn{eqn:lensing_op} is simply the integrand of the 1st order deflection angle when solving the geodesic equation (see \eqn{eqn:1storder_def} in the next section), justifying its moniker. In other words, $\opb{\tilde{r}}{a}\chi(r{\nh})$ describes the lensing action of a lens at $\tilde{r}$ on an object $\chi$ at $r$, which can be a lens $\partial^a\Psi_W$ or the source itself $\Theta^{[I]}$.

Crucially, the $N$-th order lensing operator acts linearly on $\Theta^{[N]}$, which is a nested integral of $\Theta^{[N-1]}$, $\Theta^{[N-2]}$ etc, until it terminates at $\Theta^{[I]}$ which itself is an integral over the source $S_T$ (e.g. \eqn{eqn:lensing_source}). At each order, \eqn{eqn:BE_4} tells us that each nested integral includes a Weyl term $\opa{a}\Psi_W(\bf x)$ -- this is simply a lens at location ${\bf x}$. Thus \emph{the $N$-th order lensing operator acts on all lenses associated with terms of order $N-1$ and below}, until it terminates with a lensing of the source itself. The linearity of the operator allows us to use the product rule to generate all possible lensing terms, with the proper time ordering automatically enforced by the nested integral structure -- a near lens will lense all far lenses and the source, but not \emph{vice versa}. At $N$-th order, $\Theta^{[N]}$ will have $(N-1)!$ distinct terms. 

Schematically, 
\begin{eqnarray}
{\Theta}^{[N+1]}(\hat{\mathbf{n}}) &=& \int^{\eta_0}_{\eta_{s}} d \eta_{N} \hspace{1mm} {\nabla}^{a_N}_{\hat{\mathbf{n}}} 
\left[{\Psi}_{W} (r_{N}) \right] 
({\hat{\Box}}_{\hat{\mathbf{n}}, r_N})_{a_N} \bigg\{ ... \nonumber \\
& & \int^{\eta_3}_{\eta_{s}} d \eta_2	 
{\nabla}^{a_2}_{\hat{\mathbf{n}}} \left[ {\Psi}_{W} (r_{2}) \right] 
({\hat{\Box}}_{\hat{\mathbf{n}}, r_2})_{a_2} \bigg\{  \nonumber \\
& & \int^{\eta_2}_{\eta_{s}} d \eta_1	 
{\nabla}^{a_1}_{\hat{\mathbf{n}}} \left[ {\Psi}_{W} (r_{1}) \right] 
({\hat{\Box}}_{\hat{\mathbf{n}}, r_1})_{a_1} \Theta(\hat{\mathbf{n}}) \bigg\} ... \bigg\} .
\end{eqnarray}

As shown in \cite{Su:2014mga}, ignoring the action of the lensing operator $\opb{r}{a}$ on all $\Psi_W$ terms except the innermost term in the nested integral is equivalent to ignoring all lens-lens couplings. Restoring lens-lens couplings, we will now show that \emph{all} the terms of the remapping \emph{ansatz} are generated.

\section{Equivalence of the Remapping and Boltzmann Methods} \label{equivalence}

The remapping approach \cite{Lewis:2006fu} begins with the following \emph{ansatz}, where the lensed temperature anisotropy $\tilde{\Theta}(\nh) \equiv \Theta(\nh+\hat{\alpha})$ is expressed as a Taylor expansion of its unlensed counterpart $\Theta(\nh)$, in terms of the \emph{total deflection angle} $\alpha^a = \alpha^{a}{}^{[I]}+\alpha^{a}{}^{[II]} + \dots$, with the observer plane at $r=0$,
\begin{eqnarray}
\Theta(\nh + \na)& \approx& \Theta(\nh) + \alpha^a\opc{a}\tn + \frac{1}{2}\alpha^a\alpha^b\opc{a}\opc{b}\tn + \dots \nn
&\equiv&\Theta^{[I]}(\nh) + \Theta^{[II]}(\nh)+\Theta^{[III]}(\nh)+\dots.\label{eqn:theta_expansion}
\end{eqnarray}
where $\opc{a} \equiv \partial/\partial n^{a}$ and the index $a=1,2,3$ runs over the spatial components of the coordinate basis. These terms are identified, order by order, to the corresponding higher order temperature anisotropies $\Theta^{(N)}(\nh)$.

The intervening metric perturbations -- namely the Weyl potential $\Psi_W$ -- between the source plane at $r_s$ and the observer plane at $r=0$ are encoded in the deflection $\delta \bx$ around the unlensed path $\bar{\bx}$, which is obtained by solving the geodesic equation
\begin{equation}
\frac{\partial^2 \delta x^a}{\partial r^2} = - 2\partial^a \Psi_W(\bar{\bx} + \delta \bx)~, \label{eqn:geodesic_equation}
\end{equation}
where $\partial^a \equiv \delta^{ab}\partial/\partial x^b$.  To find the deflection at each order, we solve \eqn{eqn:geodesic_equation} order by order by expanding $\delta \bx =  \delta \bx^{[I]} + \delta \bx^{[II]} + \dots$, and the Weyl potential $\Psi_W({\bar{\bx}}+\delta \bx) = \Psi_W(\bar{\bx}) + \partial_a\Psi_W(\bar{\bx})\delta x^a + (1/2)\partial_a\partial_b\Psi_W(\bar{\bx})\delta x^a\delta x^b +\dots$.
At first order, using the standard Green's function technique, the solution is
\begin{equation}
\delta x^a{}^{[I]}(r\nh) = -2r\int_{0}^{r} \dd r_1~ W(r_1,r) r_1 \partial^a\Psi_W(r_1\nh)~, \label{eqn:1storder_def}
\end{equation}
recalling that $r\nh =  \bar{\bx}$. Henceforth, we drop $\nh$ from the argument for simplicity -- it is implicitly assumed that all integrands are evaluated at the background path $\bar{\bx}$. 
The deflection angle is given by the small angle formula
\begin{equation}
\alpha^a{}^{[N]}(r) = \frac{\delta x^a{}^{[N]}(r)}{r}~. \label{eqn:small_angle}
\end{equation}
Using the remapping series expansion (\ref{eqn:theta_expansion}), the first order temperature anisotropies, \({\Theta}^{[I]} \), are unlensed. Lensing appears at second order in the expansion
\begin{eqnarray}
{\Theta}^{[II]} (\hat{\mathbf{n}}) &=&  {\alpha}^{a [I]}  \opc{a} \Theta(\hat{\mathbf{n}}) \nonumber \\
&=& -2 \int^{r_{s}}_{0} \mathrm{d} r_1 \hspace{1mm} W(r_1, r_{s})  \opa{a}{\Psi}_{W}(r_1) \opc{a} \Theta(\hat{\mathbf{n}}),
\end{eqnarray}
where we have defined the position of the source to be \(r_s\), the conformal distance from the observer to last scattering surface of the CMB. It can be seen from the form of the integral, which is first order in \(\Psi_{W}\) that this represents the anisotropies being lensed a single time between the source and the observer. 
This methodology is standard \cite{Lewis:2006fu}, and terms up to $\alpha^{[III]}$ and $\Theta^{[IV]}$ have been calculated in the literature \cite{Pratten:2016dsm,Marozzi:2016uob,Fabbian:2017wfp,Krause:2009yr,Calabrese:2014gla}.
In this work, we introduce a diagrammatic approach (detailed in Appendix \ref{sect:diagramrules}) in order to make the time ordering structure of the solutions manifest\footnote{Note that this diagrammatic formalism is not equivalent to the diagrammatic method introduced for the Boltzmann method in \cite{Su:2014mga}.}.  Using \eqn{eqn:1storder_def}, we define
\begin{equation}
\delta x^a{}^{[I]}(r) \equiv
\begin{tikzpicture}[baseline=(current bounding box.west)]
\fill (0,0) circle[radius=\dotsize]; \draw (0,0) node[below]{$r$};
\fill (1.0,0) \Square{\dotsize}; \draw (1.0,0) node[below]{$r_1$};
\draw[dashed] (0,0) to [out=60, in=120] (1.0,0);
\end{tikzpicture}
\label{eqn:deltaI}
\end{equation}
where we diagrammatically illustrate that the source at $r$ is being lensed by a lens at $r_1$ -- setting $r$ equal to $r_s$ recovers the first order deflection. Operationally,  one can read the dashed line as an integral over the kernel $-2\int_0^rdr_1 W(r,r_1)r_1\partial^a\chi(r_1)$ acting on $\chi(r_1)$ which is one of the terms of the Taylor expansion of $\Psi_W(\bar{x}^a + \delta x^a)$ i.e $\Psi_W(r)$, $\partial_b\Psi_W\delta x^b$ and   $(1/2)\partial_b\partial_c\Psi_W \delta x^b\delta x^c + \dots$ etc. \emph{Crucially, the arguments of the kernel $W(r,r_1)$ must be the upper limit of the integral $r$ and the position $r_1$ of $\chi(r_1)$ which is integrated over.} This is a direct consequence of solving the geodesic equation using the Green's function. Since the geodesic  equation is a 2nd order linear evolution equation, standard perturbation theory tells us that the order $N$-th term is a nested time integral over all the possible permutations of lower order terms which summed to $N-1$. This nested time integrals encodes a strict time ordering as follows.  At second order (see e.g. \cite{Marozzi:2016uob})
\begin{eqnarray}
\delta x^a{}^{[II]}(r) &=& -2\int^r_0 dr_1 W(r_1,r)r_1\partial^a\partial_b\Psi_W(r_1)\underbrace{\delta x^b{}^{[I]}(r_1)}_{
\begin{tikzpicture}[baseline=(current bounding box.west)]
\fill (0,0) circle[radius=\dotsize]; \draw (0,0) node[below]{$r_1$};
\fill (0.6,0) \Square{\dotsize}; \draw (0.6,0) node[below]{$r_2$};
\draw[dashed] (0,0) to [out=60, in=120] (0.6,0);
\end{tikzpicture}
} \nn
&=&
\begin{tikzpicture}[baseline=(current bounding box.west)]
\fill (0,0) circle[radius=\dotsize]; \draw (0,0) node[below]{$r$};
\fill (1.,0) \Square{\dotsize}; \draw (1.,0) node[below]{$r_1$};
\fill (2.,0) \Square{\dotsize}; \draw (2.,0) node[below]{$r_2$};
\draw[dashed] (0,0) to [out=60, in=120] (1.,0);
\draw[dashed] (1.,0) to [out=60, in=120] (2.,0);
\end{tikzpicture}
\label{eqn:deltaII}
\end{eqnarray}
which intuitively shows that the lens at $r_2$ lenses a lens at $r_1$ which itself lenses the source (or possibly another lens) at $r$. We can use these archetypes to construct higher order terms iteratively, for example at third order
\begin{widetext}
\begin{align}
\delta x^{a [III]}(r) &= -2r \int^{r}_{0} \dd r_1 \hspace{1mm} W(r_1, r) r_1 \partial^{a}
[ \partial_b \Psi_{W} (r_1) \underbrace{\delta x^{b [II]}(r_1)}_{
\begin{tikzpicture}[baseline=(current bounding box.west)]
\fill (0,0) circle[radius=\dotsize]; \draw (0,0) node[below]{$r_1$};
\fill (0.6,0) \Square{\dotsize}; \draw (0.6,0) node[below]{$r_2$};
\fill (1.2,0) \Square{\dotsize}; \draw (1.2,0) node[below]{$r_3$};
\draw[dashed] (0,0) to [out=60, in=120] (0.6,0);
\draw[dashed] (0.6,0) to [out=60, in=120] (1.2,0);
\end{tikzpicture}}
+ {1 \over 2} \partial_c \partial_b \Psi_{W} (r_1) \underbrace{\delta x^{b [I]} (r_1)}_{
\begin{tikzpicture}[baseline=(current bounding box.west)]
\fill (0,0) circle[radius=\dotsize]; \draw (0,0) node[below]{$r_1$};
\fill (0.6,0) \Square{\dotsize}; \draw (0.6,0) node[below]{$r_2$};
\draw[dashed] (0,0) to [out=60, in=120] (0.6,0);
\end{tikzpicture}
}
\underbrace{\delta x^{c [I]} (r_1)}_{
\begin{tikzpicture}[baseline=(current bounding box.west)]
\fill (0,0) circle[radius=\dotsize]; \draw (0,0) node[below]{$r_1$};
\fill (0.6,0) \Square{\dotsize}; \draw (0.6,0) node[below]{$r_3$};
\draw[dashed] (0,0) to [out=60, in=120] (0.6,0);
\end{tikzpicture}
}] \nn
&= 
\begin{tikzpicture}[baseline=(current bounding box.west)]
\fill (0,0) circle[radius=\dotsize]; \draw (0,0) node[below]{$r$};
\fill (0.6,0) \Square{\dotsize}; \draw (0.6,0) node[below]{$r_1$};
\fill (1.2,0) \Square{\dotsize}; \draw (1.2,0) node[below]{$r_2$};
\fill (1.8,0) \Square{\dotsize}; \draw (1.8,0) node[below]{$r_3$};
\draw[dashed] (0,0) to [out=60, in=120] (0.6,0);
\draw[dashed] (0.6,0) to [out=60, in=120] (1.2,0);
\draw[dashed] (1.2,0) to [out=60, in=120] (1.8,0);
\end{tikzpicture}
+ 
\begin{tikzpicture}[baseline=(current bounding box.west)]
\fill (0,0) circle[radius=\dotsize]; \draw (0,0) node[below]{$r$};
\fill (0.6,0) \Square{\dotsize}; \draw (0.6,0) node[below]{$r_1$};
\fill (1.2,0) \Square{\dotsize}; \draw (1.2,0) node[below]{$r_2$};
\draw (1.8,0)  node{$\times$};
\fill (2.4,0) \Square{\dotsize}; \draw (2.4,0) node[below]{$r_1$};
\fill (3.0,0) \Square{\dotsize}; \draw (3.0,0) node[below]{$r_3$};
\draw[dashed] (0,0) to [out=60, in=120] (0.6,0);
\draw[dashed] (0.6,0) to [out=60, in=120] (1.2,0);
\draw[dashed] (2.4,0) to [out=60, in=120] (3.0,0);
\draw[dashed] (0,0) to [out=60, in=120] (2.4,0);
\end{tikzpicture} 
\equiv
\begin{tikzpicture}[baseline=(current bounding box.west)]
\fill (0,0) circle[radius=\dotsize]; \draw (0,0) node[below]{$r$};
\fill (0.6,0) \Square{\dotsize}; \draw (0.6,0) node[below]{$r_1$};
\fill (1.2,0) \Square{\dotsize}; \draw (1.2,0) node[below]{$r_2$};
\fill (1.8,0) \Square{\dotsize}; \draw (1.8,0) node[below]{$r_3$};
\draw[dashed] (0,0) to [out=60, in=120] (0.6,0);
\draw[dashed] (0.6,0) to [out=60, in=120] (1.2,0);
\draw[dashed] (1.2,0) to [out=60, in=120] (1.8,0);
\end{tikzpicture}
+
\begin{tikzpicture}[baseline=(current bounding box.west)]
\fill (0,0) circle[radius=\dotsize]; \draw (0,0) node[below]{$r$};
\fill (0.6,0) \Square{\dotsize}; \draw (0.6,0) node[below]{$r_1$};
\fill (1.2,0) \Square{\dotsize}; \draw (1.2,0) node[below]{$r_2$};
\fill (1.8,0) \Square{\dotsize}; \draw (1.8,0) node[below]{$r_3$};
\draw[dashed] (0,0) to [out=60, in=120] (0.6,0);
\draw[dashed] (0.6,0) to [out=60, in=120] (1.2,0);
\draw[dashed] (0.6,0) to [out=60, in=120] (1.8,0);
\end{tikzpicture}
\label{eqn:deltaIII}
\end{align}
\end{widetext}
where in the last line we have combined the diagrams to make the  integral structure manifest -- lenses at $r_2$ and $r_3$ lense the lens at $r_1$, which itself lenses the source at $r$. Note that the term is symmetric under $r_2 \leftrightarrow r_3$. 

We can use these archetypes to calculate individual contributions to $\Theta(\nh)^{[N]}$. The third order anisotropies are given by
\begin{eqnarray}\label{remapping third order}
{\Theta}^{[III]}(\hat{\mathbf{n}}) = \alpha^{a[II]}  \opc{a} \Theta(\hat{\mathbf{n}}) + {1\over 2} \alpha^{a[I]} \alpha^{b[I]}
\opc{a} \opc{b} \Theta(\hat{\mathbf{n}}) 
\end{eqnarray}
The first term is given by \eqn{eqn:deltaII}, whilst the second is the product of a pair of \(\delta x^{a}{}^{[I]}\) terms, so that we have the following pair of diagrams
\begin{eqnarray}
r_s\alpha^{b[II]} = 
\begin{tikzpicture}[baseline=(current bounding box.west)]
\fill (0,0) circle[radius=\dotsize]; \draw (0,0) node[below]{$r_s$};
\fill (0.6,0) \Square{\dotsize}; \draw (0.6,0) node[below]{$r_1$};
\fill (1.2,0) \Square{\dotsize}; \draw (1.2,0) node[below]{$r_2$};
\draw[dashed] (0,0) to [out=60, in=120] (0.6,0);
\draw[dashed] (0.6,0) to [out=60, in=120] (1.2,0);
\end{tikzpicture} ~, \label{eqn:deltaIdeltaII}\nn
\end{eqnarray}
\begin{eqnarray}
r_s^2{\alpha}^{a [I]} {\alpha}^{b [I]}  &=&
\begin{tikzpicture}[baseline=(current bounding box.west)]
\fill (0,0) circle[radius=\dotsize]; \draw (0,0) node[below]{$r_s$};
\fill (0.6,0) \Square{\dotsize}; \draw (0.6,0) node[below]{$r_1$};
\draw[dashed] (0,0) to [out=60, in=120] (0.6,0);
\end{tikzpicture}
\times
\begin{tikzpicture}[baseline=(current bounding box.west)]
\fill (0,0) circle[radius=\dotsize]; \draw (0,0) node[below]{$r_s$};
\fill (0.6,0) \Square{\dotsize}; \draw (0.6,0) node[below]{$r_2$};
\draw[dashed] (0,0) to [out=60, in=120] (0.6,0);
\end{tikzpicture}\nn
&=&
\begin{tikzpicture}[baseline=(current bounding box.west)]
\fill (0,0) circle[radius=\dotsize]; \draw (0,0) node[below]{$r_s$};
\fill (0.6,0) \Square{\dotsize}; \draw (0.6,0) node[below]{$r_1$};
\fill (1.2,0) \Square{\dotsize}; \draw (1.2,0) node[below]{$r_2$};
\draw[dashed] (0.0,0) to [out=60, in=120] (0.6,0);
\draw[dashed] (0.0,0) to [out=60, in=120] (1.2,0);
\end{tikzpicture} \label{eqn:alphaI_cube}~.
\end{eqnarray}
The third order anisotropies are then
\begin{eqnarray}\label{Pratten third order}
{\Theta}^{[III]}(\hat{\mathbf{n}}) 
&=& 4 \int^{r_{s}}_{0} \hspace{-1mm} d r_1 \hspace{0.5mm} W(r_1, r_{s})  \int^{r_1}_{0} \hspace{-1mm} d r_2 \hspace{0.5mm} W(r_2, r_1) 
\hspace{-1mm} \left[ \opa{a}\opc{b}{\Psi}_{W}(r_1) \opa{b}{\Psi}_{W}(r_2) \right]
\hspace{-1mm} \opc{a} \Theta(\hat{\mathbf{n}}) \nonumber \\
&+& 2 \int^{r_{s}}_{0} \hspace{-1mm} d r_1 \hspace{0.5mm} W(r_1, r_{s})  \hspace{0.5mm} \opa{a}{\Psi}_{W}(r_1) 
\int^{r_{s}}_{0} \hspace{-1mm} d r_2 \hspace{0.5mm} W(r_2, r_{s}) 
  \opa{b}{\Psi}_{W}(r_2)  \opc{a} \opc{b} \Theta(\hat{\mathbf{n}}) . \nonumber \\
\end{eqnarray}
The second order ${\cal O}(\Psi_{W})^2$ terms evidence the fact that this represents two lensing events between the source and the observer. The second term simply represents the anisotropies being lensed a pair of times, whereas the first term is due to our earlier perturbative expansions of the Weyl potential and transverse deflection. It represents lens-lens coupling, i.e. the distortion of the first lens by the second. 

At 4th order, $\Theta(\nh)^{[IV]}$ must contain various permutations of $\alpha^{a}{}^{[I]}$, $\alpha^{a}{}^{[II]}$ and $\alpha^{a}{}^{[III]}$ terms \cite{Pratten:2016dsm,Marozzi:2016uob} 
\begin{widetext}
\begin{align}
{\Theta}^{[IV]}(\hat{\mathbf{n}})  = 
{\alpha}^{a [III]} \opc{a}\Theta(\hat{\mathbf{n}})
+   {1\over 2}\left({\alpha}^{a [I]} {\alpha}^{b [II]} + {\alpha}^{a [II]} {\alpha}^{b [I]}\right)
 \opc{a}\opc{b}\Theta(\hat{\mathbf{n}}) 
+ {1\over 3!}{\alpha}^{a [I]} {\alpha}^{b [I]} {\alpha}^{c [I]} \opc{a}\opc{b}\opc{c}\Theta(\hat{\mathbf{n}}) ~.\nn \label{eqn:4thorderterm_intro}
\end{align}
\end{widetext}
These terms are well known in the literature (e.g. \cite{Pratten:2016dsm,Marozzi:2016uob,Krause:2009yr}) -- in terms of our diagrammatic formalism they are as follows.  

The  first term of \eqn{eqn:4thorderterm_intro} is directly obtained from \eqn{eqn:deltaIII} and \eqn{eqn:small_angle}, with $r=r_s$ and noting that the $\partial_a\Theta(\nh)$ term can be brought into the integral since it is not a function of time.  This gives us two total terms which are
\begin{widetext}
\begin{align}
\Theta^{IV}&\supset  \frac{\delta x^{a}{}^{[III]}}{r}\opc{a} \Theta =
\begin{tikzpicture}[baseline=(current bounding box.west)]
\fill (0,0) circle[radius=\dotsize]; \draw (0,0) node[below]{$r$};
\fill (0.6,0) \Square{\dotsize}; \draw (0.6,0) node[below]{$r_1$};
\fill (1.2,0) \Square{\dotsize}; \draw (1.2,0) node[below]{$r_2$};
\fill (1.8,0) \Square{\dotsize}; \draw (1.8,0) node[below]{$r_3$};
\draw[dashed] (0,0) to [out=60, in=120] (0.6,0);
\draw[dashed] (0.6,0) to [out=60, in=120] (1.2,0);
\draw[dashed] (1.2,0) to [out=60, in=120] (1.8,0);
\end{tikzpicture} 
+
\begin{tikzpicture}[baseline=(current bounding box.west)]
\fill (0,0) circle[radius=\dotsize]; \draw (0,0) node[below]{$r$};
\fill (0.6,0) \Square{\dotsize}; \draw (0.6,0) node[below]{$r_1$};
\fill (1.2,0) \Square{\dotsize}; \draw (1.2,0) node[below]{$r_2$};
\fill (1.8,0) \Square{\dotsize}; \draw (1.8,0) node[below]{$r_3$};
\draw[dashed] (0,0) to [out=60, in=120] (0.6,0);
\draw[dashed] (0.6,0) to [out=60, in=120] (1.2,0);
\draw[dashed] (0.6,0) to [out=60, in=120] (1.8,0);
\end{tikzpicture}
\end{align}
\end{widetext}

The second term is the product of $\delta x^{a}{}^{[I]}$ and $\delta x^{a}{}^{[II]}$, which can be constructed as
\begin{eqnarray}
r_s^2\alpha^{a[I]}\alpha^{b[II]} = 
\begin{tikzpicture}[baseline=(current bounding box.west)]
\fill (0,0) circle[radius=\dotsize]; \draw (0,0) node[below]{$r_s$};
\fill (0.6,0) \Square{\dotsize}; \draw (0.6,0) node[below]{$r_1$};
\draw[dashed] (0,0) to [out=60, in=120] (0.6,0);
\end{tikzpicture}
\times
\begin{tikzpicture}[baseline=(current bounding box.west)]
\fill (0,0) circle[radius=\dotsize]; \draw (0,0) node[below]{$r_s$};
\fill (0.6,0) \Square{\dotsize}; \draw (0.6,0) node[below]{$r_2$};
\fill (1.2,0) \Square{\dotsize}; \draw (1.2,0) node[below]{$r_3$};
\draw[dashed] (0,0) to [out=60, in=120] (0.6,0);
\draw[dashed] (0.6,0) to [out=60, in=120] (1.2,0);
\end{tikzpicture} = 
\begin{tikzpicture}[baseline=(current bounding box.west)]
\fill (0,0) circle[radius=\dotsize]; \draw (0,0) node[below]{$r_s$};
\fill (0.6,0) \Square{\dotsize}; \draw (0.6,0) node[below]{$r_1$};
\fill (1.2,0) \Square{\dotsize}; \draw (1.2,0) node[below]{$r_2$};
\fill (1.8,0) \Square{\dotsize}; \draw (1.8,0) node[below]{$r_3$};
\draw[dashed] (0,0) to [out=60, in=120] (0.6,0);
\draw[dashed] (0.0,0) to [out=60, in=120] (1.2,0);
\draw[dashed] (1.2,0) to [out=60, in=120] (1.8,0);
\end{tikzpicture} ~. \nn
\end{eqnarray}
Physically this captures the lensing of the source by a lens at $r_1$, while the lenses at $r_3$ lenses another lens at $r_2$ which then lenses the source. 
Its symmetric counterpart $\alpha^{a[II]}\alpha^{b[I]}$ is obtained under the permutation $r_1\rightarrow r_3$, $r_2\rightarrow r_1$ and $r_3\rightarrow r_2$.
The third term in our diagrammatic method is
\begin{eqnarray}
r_s^3{\alpha}^{a [I]} {\alpha}^{b [I]} {\alpha}^{c [I]} &=&
\begin{tikzpicture}[baseline=(current bounding box.west)]
\fill (0,0) circle[radius=\dotsize]; \draw (0,0) node[below]{$r_s$};
\fill (0.6,0) \Square{\dotsize}; \draw (0.6,0) node[below]{$r_1$};
\draw[dashed] (0,0) to [out=60, in=120] (0.6,0);
\end{tikzpicture}
\times
\begin{tikzpicture}[baseline=(current bounding box.west)]
\fill (0,0) circle[radius=\dotsize]; \draw (0,0) node[below]{$r_s$};
\fill (0.6,0) \Square{\dotsize}; \draw (0.6,0) node[below]{$r_2$};
\draw[dashed] (0,0) to [out=60, in=120] (0.6,0);
\end{tikzpicture}
\times
\begin{tikzpicture}[baseline=(current bounding box.west)]
\fill (0,0) circle[radius=\dotsize]; \draw (0,0) node[below]{$r_s$};
\fill (0.6,0) \Square{\dotsize}; \draw (0.6,0) node[below]{$r_3$};
\draw[dashed] (0,0) to [out=60, in=120] (0.6,0);
\end{tikzpicture} \nn
&=&
\begin{tikzpicture}[baseline=(current bounding box.west)]
\fill (0,0) circle[radius=\dotsize]; \draw (0,0) node[below]{$r_s$};
\fill (0.6,0) \Square{\dotsize}; \draw (0.6,0) node[below]{$r_1$};
\fill (1.2,0) \Square{\dotsize}; \draw (1.2,0) node[below]{$r_2$};
\fill (1.8,0) \Square{\dotsize}; \draw (1.8,0) node[below]{$r_3$};
\draw[dashed] (0,0) to [out=60, in=120] (1.8,0);
\draw[dashed] (0.0,0) to [out=60, in=120] (0.6,0);
\draw[dashed] (0.0,0) to [out=60, in=120] (1.2,0);
\end{tikzpicture}~,
\end{eqnarray}
with the physical interpretation that the source at $r_s$ is lensed separately by lenses at $r_1$, $r_2$ and $r_3$ -- i.e. this is the sole term if we ignore all lens-lens couplings.
As we have mentioned in section \ref{sect:boltzmann}, while these terms are often called ``post-Born'' in the literature (e.g. \cite{Petri:2016qya,Barthelemy:2020igw}), the Weyl potentials $\Psi_W$ in the integrands are evaluated along the unperturbed path $\bar{\bx}$, and hence morally speaking we are still applying the Born Approximation.

Summing these terms together, at fourth order the anisotropies are 
\begin{eqnarray}\label{Pratten fourth order}
{\Theta}^{[IV]}_{remapping} (\hat{\mathbf{n}}) = A' + B' + C' + D' + F',
\end{eqnarray}
where
\begin{eqnarray}
A' &=&  - \Bigg\{ 8 \int^{r_{s}}_{0} \hspace{-1mm} d r_1 \hspace{1mm} W(r_1, r_{s})  
\hspace{-1mm} \int^{r_1}_{0} \hspace{-1mm} d r_2 \hspace{0.5mm} W(r_2, r_1)  \nonumber \\
& & \hspace{-3mm} \int^{r_2}_{0} \hspace{-1.5mm} 
d r_3 \hspace{1mm} W(r_3, r_2) 
\big[\opa{a}\opc{b}{\Psi}_{W}(r_1) \opa{b}\opc{c}{\Psi}_{W}(r_2)\opa{c}{\Psi}_{W}(r_3) \big] 
\opc{a} \Theta(\hat{\mathbf{n}}) \hspace{-1mm} \Bigg\} , \label{first term} \\ 
B'&=& \hspace{-1mm} -  \Bigg\{ \hspace{-0.5mm} 4 \int^{r_{s}}_{0} \hspace{-1mm} d r_1 \hspace{0.5mm} W(r_1, r_{s})  
\hspace{-1mm} \int^{r_1}_{0} \hspace{-1mm} d r_2 \hspace{0.5mm} W(r_2, r_1)  \big[\opa{a}\opc{b}\opc{c}{\Psi}_{W}(r_1) \opa{b}{\Psi}_{W}(r_2) \big] \nonumber \\ 
& & \int^{r_1}_{0} \hspace{-1mm} d r_3 \hspace{1mm} W(r_3, r_1) 
\opa{c}{\Psi}_{W}(r_3) 
\opc{a} \Theta(\hat{\mathbf{n}}) \Bigg\} , \label{second term} \\ 
C'&=& - \Bigg\{4 \bigg[ \int^{r_{s}}_{0} \hspace{-1mm} d r_1 \hspace{1mm} W(r_1, r_{s})  \opa{a}{\Psi}_{W}(r_1) \bigg]
\hspace{-1mm} \bigg[ \int^{r_{s}}_{0} \hspace{-1mm} d r_2 \hspace{1mm} W(r_2, r_{s})  \nonumber \\
& & \int^{r_2}_{0} \hspace{-1mm} d r_3 \hspace{1mm} W(r_3, r_2) 
 \opa{b}\opc{c}{\Psi}_{W}(r_2)\opa{c}{\Psi}_{W}(r_3) \bigg]
\opc{a} \opa{b} \Theta(\hat{\mathbf{n}}) \Bigg\} , \label{third term} \\ \nonumber
D'&=& - \Bigg\{ 4 \bigg[ \int^{r_{s}}_{0} \hspace{-1mm} d r_1 \hspace{1mm} W(r_1, r_{s})   
\hspace{-1mm} \int^{r_1}_{0} \hspace{-1mm} d r_2 \hspace{1mm} W(r_2, r_1)  \opa{a}\opc{c}{\Psi}_{W}(r_1) \opa{c}{\Psi}_{W}(r_2) \bigg]
\nonumber \\
& & \bigg[ \int^{r_{s}}_{0} \hspace{-1mm} d r_3 \hspace{1mm} W(r_3, r_{s})  
\opa{b}{\Psi}_{W}(r_3) \bigg]
\opc{a} \opc{b} \Theta(\hat{\mathbf{n}}) \Bigg\} , \label{fourth term} \\ 
F'&=& - \Bigg\{ {4\over 3}  \bigg[ \int^{r_{s}}_{0} \hspace{-2mm} d r_1 \hspace{1mm} W(r_1, r_{s})  \opa{a}{\Psi}_{W}(r_1) \bigg]
 \hspace{-1mm} \bigg[ \int^{r_{s}}_{0} \hspace{-2mm} d r_2 \hspace{1mm} W(r_2, r_{s})  \opa{b}{\Psi}_{W}(r_2) \bigg] \nonumber \\
& & \bigg[ \int^{r_{s}}_{0} \hspace{-2mm} d r_3 \hspace{1mm} W(r_3, r_{s})  \opa{c}{\Psi}_{W}(r_3) \bigg]
\opc{a} \opc{b} \opc{c} \Theta(\hat{\mathbf{n}}) \Bigg\} . \label{fifth term} 
\end{eqnarray}
Note that there is no term \(E'\) as an additional term \(E\) initially appears to be unique to the Boltzmann equation solution as we will see later; the terms are ordered by the number of couplings to the source. At $\Theta^{[IV]}(\nh)$ order, there are a total of 5 terms, although the symmetric \eqn{eqn:deltaIdeltaII} terms are counted twice \cite{Krause:2009yr}. The number of terms, including symmetric ones, in such an expansion is given by the Catalan number $C_{N-1}$ with  $C_N=\frac{1}{N+1}\binom{2N}{N}$.  So at $\Theta^{[V]}$ order, there are 14 terms etc. We prove this in Appendix \ref{sect:diagramrules}.

\subsection{Equivalence at Third Order} \label{eato}

As a warm up to the fourth order problem, we will now prove the equivalence of the Boltzmann method with the canonical remapping method at third order. Using the Boltzmann formalism, we find at this order
\begin{eqnarray}\label{third order result}
{\Theta}^{[III]}(\hat{\mathbf{n}}) &=& 4\int^{r_{s}}_{0} d r_{2} \hspace{1mm} \opa{a} 
{\Psi}_{W} (r_{2})  \bigg\{ \nonumber \\ \nonumber
& & \hspace{-10mm} \int^{r_{s}}_{r_2} d r_{1} W(r_1 , r_2) W(r_s , r_1) 
\opa{b} \opc{a} {\Psi}_{W} (r_{1})  
\opc{b} \Theta(\hat{\mathbf{n}}) \\ \nonumber
& & \hspace{-10mm} + \int^{r_{s}}_{r_2} d r_{1} W(r_s , r_2) W(r_s , r_1)
\opa{b} {\Psi}_{W} (r_{1})  
\opc{a} \opc{b} \Theta(\hat{\mathbf{n}}) \bigg\} . \\
\end{eqnarray}
As we will prove below, this equation \eqn{third order result} is completely equivalent to \eqn{Pratten third order} derived using the canonical method; it represents two lensing events between the source and the observer. The second term  represents the anisotropies being lensed a pair of times, the first  represents lens-lens coupling. In comparison with the geodesic equation method, we did not have to relax the Born approximation to include \textquotedblleft ray-deflection\textquotedblright terms to produce this result. The above formula is due to gravitational lensing only, with no redshift, time-delay or ray-deflection. 
That we have been able to derive \eqn{Pratten third order} implies that the the ray-deflection terms included in previous work \cite{Shapiro:2006em}\cite{Pratten:2016dsm} are in fact also due to lens-lens coupling. In contrast with the previous method, solving for each order in a consistent way, rather than dealing with multiple Taylor expansions, has resulted in a single integration path for both terms.

We will now show that \eqn{third order result} is equivalent to \eqn{Pratten third order} by first rearranging the order of the integration limits using the formula
\begin{eqnarray}\label{rearrange order}
\int^{r_{M+1}}_{0} d r_{M} \int^{r_{M}}_{0} d \tilde{r} = \int^{r_{M+1}}_{0} d \tilde{r} \int^{r_{M+1}}_{\tilde{r}} d r_{M}
\end{eqnarray}
to reverse the integration path of \eqn{third order result} and obtain
\begin{eqnarray}
{\Theta}^{[III]}(\hat{\mathbf{n}}) &=& 4 \int^{r_{s}}_{0} d r_{1} W(r_s , r_1)  \bigg\{ \nonumber \\ 
& & \hspace{-10mm} \int^{r_{1}}_{0} d r_{2} W(r_1 , r_2)   \opa{a} 
{\Psi}_{W} (r_{2})  
\opa{b} \opc{a} {\Psi}_{W} (r_{1})  
\opc{b} \Theta(\hat{\mathbf{n}}) \bigg\} \nonumber \\ 
&+& 4 \int^{r_{s}}_{0} d r_{1} W(r_s , r_1) 
\opa{b} {\Psi}_{W} (r_{1})  \bigg\{ \nonumber \\ 
& & \int^{r_{1}}_{0} d r_{2} W(r_s , r_2)  
\opa{a} {\Psi}_{W} (r_{2}) 
\opc{a}\opc{b} \Theta(\hat{\mathbf{n}}) \bigg\} .
\end{eqnarray}
The second term has a symmetry that allows the integrals to be decoupled using the relation
\begin{eqnarray}\label{decouple}
\int^{r_{s}}_{0} d r_{1} f(r_{1}) \int^{r_{1}}_{0} d r_{2} f(r_{2}) \cdots  \int^{r_{N-1}}_{0} d r_{N} f(r_{N})
= {1 \over N!} \int^{r_{s}}_{0} d r_{1} f(r_{1}) \int^{r_{s}}_{0} d r_{2} f(r_{2}) \cdots \int^{r_{s}}_{0} d r_{N} f(r_{N})
\end{eqnarray}
with the result being
\begin{eqnarray}
{\Theta}^{[III]}(\hat{\mathbf{n}}) &=& 4 \int^{r_{s}}_{0} d r_{1} W(r_s , r_1)  \bigg\{ \nonumber \\ 
& & \hspace{-10mm} \int^{r_{1}}_{0} d r_{2} W(r_1 , r_2)  \opa{a} {\Psi}_{W} (r_{2})  
\opa{b}
 \opc{a} {\Psi}_{W} (r_{1}) 
\opc{b} \Theta(\hat{\mathbf{n}}) \bigg\} \nonumber \\ 
&+& 2 \int^{r_{s}}_{0} d r_{1} W(r_s , r_1) 
\opa{b} {\Psi}_{W} (r_{1})  \bigg\{ \nonumber \\ 
& & \int^{r_{s}}_{0} d r_{2} W(r_s , r_2)  
\opa{a} {\Psi}_{W} (r_{2}) 
\opc{a}\opc{b} \Theta(\hat{\mathbf{n}}) \bigg\},
\end{eqnarray}
which matches the remapping method result, \eqn{Pratten third order}. The reason for the differing integration paths in results derived using the remapping method is now clear: it is because the nested integrals had already been decoupled. This is due to the multiplication of \(\alpha^{a}\) terms in \eqn{eqn:theta_expansion}. The coupling between the integrals represented by each \(\alpha^{a}\) is already broken in the remapping series. 

\subsection{Equivalence at Fourth Order} \label{eafo}

The solution from the Boltzmann equation for the fourth order anisotropies is given by the sum of the six terms, which was first derived in \cite{Su:2014mga},
\begin{eqnarray}\label{fourth order result}
{\Theta}^{[IV]}_{Boltzmann} (\hat{\mathbf{n}}) = A + B + C + D + E + F,
\end{eqnarray}
which have the form
\begin{eqnarray}
A &=& 8\int^{r_{s}}_{0} d r_{3} \hspace{1mm} \opa{c}
{\Psi}_{W} (r_{3}) ]  \Bigg\{ \nonumber \int^{r_{s}}_{r_3} d r_{2} W(r_2 , r_3) 
\opa{b} \opc{c} {\Psi}_{W} (r_{2})  
\bigg\{ \\  
& &  \int^{r_{s}}_{r_2} d r_{1} W(r_1 , r_2) W(r_s , r_1) 
\opa{a} \opc{b} {\Psi}_{W} (r_{1})  
\opc{a} \Theta(\hat{\mathbf{n}}) \bigg\} \Bigg\}, \\ \nonumber
B&=& 8\int^{r_{s}}_{0} d r_{3} \hspace{1mm} \opa{c} {\Psi}_{W} (r_{3}) 
\int^{r_{s}}_{r_3} d r_{2}  \hspace{1mm}
\opa{b}  {\Psi}_{W} (r_{2})  
\bigg\{ \nonumber \\ 
& &  \int^{r_{s}}_{r_2} d r_{1} W(r_1 , r_3) 
W(r_1 , r_2) W(r_s , r_1) \opa{a} \opc{c} \opc{b} {\Psi}_{W} (r_{1})  
\opc{a} \Theta(\hat{\mathbf{n}}) \bigg\}, \\ \nonumber
C &=& 8 \int^{r_{s}}_{0} d r_{3} \hspace{1mm} \opa{c} 
{\Psi}_{W} (r_{3})   \Bigg\{ \nonumber \int^{r_{s}}_{r_3} d r_{2}  W(r_2 , r_3)  
\opa{b} \opc{c} {\Psi}_{W} (r_{2})  
\bigg\{ \\ 
& & \int^{r_{s}}_{r_2} d r_{1} W(r_s , r_2)  W(r_s , r_1) 
\opa{a} {\Psi}_{W} (r_{1})  
\opc{b} \opc{a} \Theta(\hat{\mathbf{n}}) \bigg\} \Bigg\}, \\ \nonumber
D&=& 8\int^{r_{s}}_{0} d r_{3} \hspace{1mm} \opa{b} 
{\Psi}_{W} (r_{3})   \Bigg\{ \nonumber \int^{r_{s}}_{r_3} d r_{2}  \hspace{1mm}
\opa{c}  {\Psi}_{W} (r_{2})  
\bigg\{ \\  
& &  \int^{r_{s}}_{r_2} \hspace{-2mm} d r_{1} W(r_s , r_3) W(r_1 , r_2)  W(r_s , r_1) \opa{a}  \opc{c} {\Psi}_{W} (r_{1})  
\opc{b} \opc{a} \Theta(\hat{\mathbf{n}})  \bigg\} \Bigg\}, \\ \nonumber
E&=& 8 \int^{r_{s}}_{0} d r_{3} \hspace{1mm} \opa{c} {\Psi}_{W} (r_{3})   
\int^{r_{s}}_{r_3} d r_{2}  \hspace{1mm}
\opa{b}  {\Psi}_{W} (r_{2})  
\bigg\{ \nonumber \\  
& & \int^{r_{s}}_{r_2} \hspace{-2mm} d r_{1} W(r_1 , r_3) W(r_s , r_2)  W(r_s , r_1) \opa{a} \opc{c} {\Psi}_{W} (r_{1})  
\opc{b} \opc{a} \Theta(\hat{\mathbf{n}}) \bigg\}, \\ \nonumber
F&=& 8\int^{r_{s}}_{0} d r_{3} \hspace{1mm} \opa{c} {\Psi}_{W} (r_{3})   
\int^{r_{s}}_{r_3} d r_{2}  \hspace{1mm}
\opa{b}  {\Psi}_{W} (r_{2})  
\bigg\{ \nonumber \\  
& &  \int^{r_{s}}_{r_2} \hspace{-3mm} d r_{1} W(r_s , r_3) W(r_s , r_2)  W(r_s , r_1) \opa{a} {\Psi}_{W} (r_{1})  
\opc{c} \opc{b} \opc{a} \Theta(\hat{\mathbf{n}})  \bigg\} . 
\end{eqnarray}

As with \eqn{Pratten fourth order}, this represents three lensing events between the source and the observer. The final term represents the anisotropies being lensed a trio of times, the other terms represent various combinations of lens-lens coupling. Due to the systematic nature of this derivation we once again have a single integration path for all six terms. Unlike \eqn{Pratten fourth order}, we now have six terms, whereas the geodesic equation method yielded only five.  However, as we will show below, similar to the 3rd order case, the complicated integral structure of the two sets of terms hide their equivalence despite the differing number of terms. In other words, in comparison to the 5 terms of the canonical remapping method \eqn{Pratten fourth order}, the above terms are completely equivalent: $A=A'$, $B=B'$, $F=F'$ and $C'+D' = C+D+E$. 

Working through \eqn{fourth order result} term by term, we begin with the first term, term $A$:
\begin{eqnarray} 
A &=& 
8\int^{r_{s}}_{0} d r_{3} \hspace{1mm} \opa{c}
{\Psi}_{W} (r_{3}) ]  \Bigg\{ \nonumber \\ \nonumber
& &\int^{r_{s}}_{r_3} d r_{2} W(r_2 , r_3) 
\opa{b} \opc{c} {\Psi}_{W} (r_{2})  
\bigg\{ \\ \nonumber 
& & \hspace{-20mm} \int^{r_{s}}_{r_2} d r_{1} W(r_1 , r_2) W(r_s , r_1) 
\opa{a} \opc{b} {\Psi}_{W} (r_{1})  
\opc{a} \Theta(\hat{\mathbf{n}}) \bigg\} \Bigg\}, \\
\end{eqnarray}
and as with the third order result we use \eqn{rearrange order} to rearrange the order of integration to obtain
\begin{eqnarray} 
A &=& 
8\int^{r_{s}}_{0} d r_{1} W(r_s , r_1) 
\int^{r_{1}}_{0} d r_{2} W(r_1 , r_2) 
\int^{r_{2}}_{0} d r_{3} W(r_2 , r_3)  
\bigg\{ \nonumber \\ \nonumber 
& & \opa{c} {\Psi}_{W} (r_{3}) 
\opa{b} \opc{c} {\Psi}_{W} (r_{2})   
\opa{a} \opc{b} {\Psi}_{W} (r_{1})  
\opc{a} \Theta(\hat{\mathbf{n}}) \bigg\} , \\
\end{eqnarray}
matching the equivalent term \eqn{first term} which was one of the two terms derived from the term \({\alpha}^{a [III]} \opc{a} \Theta(\hat{\mathbf{n}})\) in the series expansion \eqn{eqn:4thorderterm_intro}. This had the diagrammatic form:
\begin{tikzpicture}[baseline=(current bounding box.west)]
\fill (0,0) circle[radius=\dotsize]; \draw (0,0) node[below]{$r$};
\fill (0.6,0) \Square{\dotsize}; \draw (0.6,0) node[below]{$r_1$};
\fill (1.2,0) \Square{\dotsize}; \draw (1.2,0) node[below]{$r_2$};
\fill (1.8,0) \Square{\dotsize}; \draw (1.8,0) node[below]{$r_3$};
\draw[dashed] (0,0) to [out=60, in=120] (0.6,0);
\draw[dashed] (0.6,0) to [out=60, in=120] (1.2,0);
\draw[dashed] (1.2,0) to [out=60, in=120] (1.8,0);
\end{tikzpicture}.
We next move on to term $B$ which also has only a single coupling to the source
\begin{eqnarray}
B &=& 
8\int^{r_{s}}_{0} d r_{3} \hspace{1mm} \opa{c} {\Psi}_{W} (r_{3}) 
\int^{r_{s}}_{r_3} d r_{2}  \hspace{1mm}
\opa{b}  {\Psi}_{W} (r_{2})  
\bigg\{ \nonumber \\ 
& &  \int^{r_{s}}_{r_2} d r_{1} W(r_1 , r_3) 
W(r_1 , r_2) W(r_s , r_1) \opa{a} \opc{c} \opc{b} {\Psi}_{W} (r_{1})  
\opc{a} \Theta(\hat{\mathbf{n}}) \bigg\}, 
\end{eqnarray}
rearranging the order of integration to obtain
\begin{eqnarray} 
B &=& 
8\int^{r_{s}}_{0} d r_{1} W(r_s , r_1) 
\opa{a} \opc{c} \opc{b} {\Psi}_{W} (r_{1})  \Bigg\{ \nonumber \\ \nonumber
& &\int^{r_{1}}_{0} d r_{2} W(r_1 , r_2) 
\opa{b}  {\Psi}_{W} (r_{2})  \bigg\{ \\  
& & \int^{r_{2}}_{0} d r_{3} W(r_1 , r_3) \opa{c} {\Psi}_{W} (r_{3}) 
\opc{a} \Theta(\hat{\mathbf{n}}) \bigg\} \Bigg\}.
\end{eqnarray}
The inner pair of integrals again have a symmetry that allows us to use \eqn{decouple} to decouple them, giving
\begin{eqnarray} 
B &=& 
4\int^{r_{s}}_{0} d r_{1} W(r_s , r_1) 
\opa{a} \opc{c} \opc{b} 
{\Psi}_{W} (r_{1})  \Bigg\{ \nonumber \\ \nonumber
& &\int^{r_{1}}_{0} d r_{2} W(r_1 , r_2)
\opa{b} {\Psi}_{W} (r_{2})  \bigg\{ \\ 
& & \int^{r_{1}}_{0} d r_{3} W(r_1 , r_3) 
\hspace{1mm} \opa{c} {\Psi}_{W} (r_{3}) 
\opc{a} \Theta(\hat{\mathbf{n}}) \bigg\} \Bigg\},
\end{eqnarray}
so that this term also now matches the equivalent term \eqn{second term}, the second of the two terms derived from the term \({\alpha}^{a [III]} \opc{a} \Theta(\hat{\mathbf{n}})\) in the series expansion \eqn{eqn:4thorderterm_intro} and has the diagrammatic form 
\begin{tikzpicture}[baseline=(current bounding box.west)]
\fill (0,0) circle[radius=\dotsize]; \draw (0,0) node[below]{$r$};
\fill (0.6,0) \Square{\dotsize}; \draw (0.6,0) node[below]{$r_1$};
\fill (1.2,0) \Square{\dotsize}; \draw (1.2,0) node[below]{$r_2$};
\fill (1.8,0) \Square{\dotsize}; \draw (1.8,0) node[below]{$r_3$};
\draw[dashed] (0,0) to [out=60, in=120] (0.6,0);
\draw[dashed] (0.6,0) to [out=60, in=120] (1.2,0);
\draw[dashed] (0.6,0) to [out=60, in=120] (1.8,0);
\end{tikzpicture}. Moving on to term $C$:
\begin{eqnarray}
C &=& 
8 \int^{r_{s}}_{0} d r_{3} \hspace{1mm} \opa{c} 
{\Psi}_{W} (r_{3})   \Bigg\{ \int^{r_{s}}_{r_3} d r_{2}  W(r_2 , r_3)  
\opa{b} \opc{c} {\Psi}_{W} (r_{2})  
\bigg\{ \nonumber \\ 
& &  \int^{r_{s}}_{r_2} d r_{1} W(r_s , r_2)  W(r_s , r_1) 
\opa{a} {\Psi}_{W} (r_{1})  
\opc{b} \opc{a} \Theta(\hat{\mathbf{n}}) \bigg\} \Bigg\}, 
\end{eqnarray}
we once more rearrange the order of integration to obtain
\begin{eqnarray}
C &=& 
8 \int^{r_{s}}_{0} d r_{1} W(r_s , r_1)  
\opa{a} {\Psi}_{W} (r_{1})  \Bigg\{ \nonumber \\ \nonumber
& &\int^{r_{1}}_{0} d r_{2} W(r_s , r_2) 
\opa{b} \opc{c} {\Psi}_{W} (r_{2}) \bigg\{ \\ 
& & \int^{r_{2}}_{0} d r_{3} W(r_2 , r_3)  
\hspace{1mm} \opa{c} {\Psi}_{W} (r_{3}) 
\opc{b} \opc{a} \Theta(\hat{\mathbf{n}}) \bigg\} \Bigg\}.
\end{eqnarray}
We repeat this process with the fourth term, $D$;
\begin{eqnarray} 
D &=& 
8\int^{r_{s}}_{0} d r_{3} \hspace{1mm} \opa{b} 
{\Psi}_{W} (r_{3})   \Bigg\{ \int^{r_{s}}_{r_3} d r_{2}  \hspace{1mm}
\opa{c}  {\Psi}_{W} (r_{2})  
\bigg\{ \nonumber \\  
& &  \int^{r_{s}}_{r_2} \hspace{-2mm} d r_{1} W(r_s , r_3) W(r_1 , r_2)  W(r_s , r_1) \opa{a}  \opc{c} {\Psi}_{W} (r_{1})  
\opc{b} \opc{a} \Theta(\hat{\mathbf{n}})  \bigg\} \Bigg\},
\end{eqnarray}
rearranging the order of integration
\begin{eqnarray} 
D &=& 
8\int^{r_{s}}_{0} d r_{1} W(r_s , r_1)
\Bigg\{ \nonumber \\ \nonumber
& & \int^{r_{1}}_{0} d r_{2} W(r_1 , r_2) 
\hspace{1mm} 
\opa{a}  \opc{c} {\Psi}_{W} (r_{1})  
\opa{c}  {\Psi}_{W} (r_{2})  
\bigg\{ \\  
& & \int^{r_{2}}_{0} \hspace{-2mm} d r_{3} W(r_s , r_3)  
\hspace{1mm} \opa{b} {\Psi}_{W} (r_{3}) 
\opc{b} \opc{a} \Theta(\hat{\mathbf{n}})  \bigg\} \Bigg\},
\end{eqnarray}
to obtain the final Boltzmann equation method result for this term. 
We now focus on the fifth term, which we label term $E$:
\begin{eqnarray} 
E &=& 
8 \int^{r_{s}}_{0} d r_{3} \hspace{1mm} \opa{c} {\Psi}_{W} (r_{3})   
\int^{r_{s}}_{r_3} d r_{2}  \hspace{1mm}
\opa{b}  {\Psi}_{W} (r_{2})  
\bigg\{ \nonumber \\  
& & \int^{r_{s}}_{r_2} \hspace{-2mm} d r_{1} W(r_1 , r_3) W(r_s , r_2)  W(r_s , r_1)  \opa{a} \opc{c} {\Psi}_{W} (r_{1})  
\opc{b} \opc{a} \Theta(\hat{\mathbf{n}}) \bigg\} ,  \nn
 &=& 
8 \int^{r_{s}}_{0} d r_{1}  W(r_s , r_1) 
\int^{r_{1}}_{0} d r_{2} W(r_s , r_2) 
\int^{r_{2}}_{0} \hspace{-2mm} d r_{3} W(r_1 , r_3) \bigg\{ \nonumber \\  
& & \opa{c} {\Psi}_{W} (r_{3}) 
\hspace{1mm} \opa{b}  {\Psi}_{W} (r_{2}) 
\hspace{1mm} \opa{a} \opc{c} {\Psi}_{W} (r_{1}) \opc{b} \opc{a} \Theta(\hat{\mathbf{n}}) \bigg\} ,
\end{eqnarray}
where we have rearranged the order of the integration as usual in the 2nd line.  The sixth term, which we shall label term $F$, is
\begin{eqnarray}
F &=&
8\int^{r_{s}}_{0} d r_{3} \hspace{1mm} \opa{c} {\Psi}_{W} (r_{3})   
\int^{r_{s}}_{r_3} d r_{2}  \hspace{1mm}
\opa{b}  {\Psi}_{W} (r_{2})  
\bigg\{ \nonumber \\ \nonumber 
& &  \int^{r_{s}}_{r_2} \hspace{-3mm} d r_{1} W(r_s , r_3) W(r_s , r_2)  W(r_s , r_1) \opa{a} {\Psi}_{W} (r_{1})  
\opc{c} \opc{b} \opc{a} \Theta(\hat{\mathbf{n}})  \bigg\} ,  \\
&=&
8\int^{r_{s}}_{0} d r_{1} W(r_s , r_1) 
\opa{a} {\Psi}_{W} (r_{1}) \Bigg\{ \nonumber \\ \nonumber
& & \int^{r_{1}}_{0} d r_{2} W(r_s , r_2) 
\hspace{1mm} \opa{b} {\Psi}_{W} (r_{2})  
\bigg\{ \\  
& &  \int^{r_{2}}_{0} \hspace{-3mm} d r_{3} W(r_s , r_3)   
\hspace{1mm} \opa{c} {\Psi}_{W} (r_{3})   
\opc{c} \opc{b} \opc{a} \Theta(\hat{\mathbf{n}})  \bigg\} \Bigg\},
\end{eqnarray}
which again we rearranged the order of the integration in the 2nd line.
This has a symmetry that allows it to be decoupled using \eqn{decouple}
after which it takes the form
\begin{eqnarray}
F &=&
{4 \over 3} \int^{r_{s}}_{0} d r_{1} W(r_s , r_1) 
\opa{a} {\Psi}_{W} (r_{1}) \Bigg\{ \nonumber \\ \nonumber
& & \int^{r_{s}}_{0} d r_{2} W(r_s , r_2) 
\hspace{1mm} \opc{b}  {\Psi}_{W} (r_{2})  \bigg\{ \\  
& &  \int^{r_{s}}_{0} \hspace{-3mm} d r_{3} W(r_s , r_3)  
\hspace{1mm} \opa{c} {\Psi}_{W} (r_{3})   
\opc{c} \opc{b} \opc{a} \Theta(\hat{\mathbf{n}})  \bigg\} \Bigg\},
\end{eqnarray}
which matches the result for the term \eqn{fifth term} produced by the product of three deflection angles; \({1\over 6}{\alpha}^{a [I]} {\alpha}^{b [I]} {\alpha}^{c [I]} \opc{a} \opc{b} \opc{c} \Theta(\hat{\mathbf{n}})\) and has the diagrammatic form 
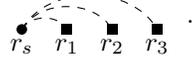
\begin{tikzpicture}[baseline=(current bounding box.west)]
\fill (0,0) circle[radius=\dotsize]; \draw (0,0) node[below]{$r_s$};
\fill (0.6,0) \Square{\dotsize}; \draw (0.6,0) node[below]{$r_1$};
\fill (1.2,0) \Square{\dotsize}; \draw (1.2,0) node[below]{$r_2$};
\fill (1.8,0) \Square{\dotsize}; \draw (1.8,0) node[below]{$r_3$};
\draw[dashed] (0,0) to [out=60, in=120] (1.8,0);
\draw[dashed] (0.0,0) to [out=60, in=120] (0.6,0);
\draw[dashed] (0.0,0) to [out=60, in=120] (1.2,0);
\end{tikzpicture} .

After these rearrangements the solution from the Boltzmann equation for the $\Theta^{(IV)}$ is now given by the sum of the six terms:
\begin{eqnarray}
{\Theta}^{[IV]}_{Boltzmann} (\hat{\mathbf{n}}) = A + B + C + D + E + F,
\end{eqnarray}
where 
\begin{eqnarray}
A &=& -  \Bigg\{ 8  \int^{r_{s}}_{0} \hspace{-1mm} d r_1 \hspace{1mm} W(r_1, r_{s})  
\hspace{-1mm} \int^{r_1}_{0} \hspace{-1mm} d r_2 \hspace{1mm} W(r_2, r_1)  \nonumber \\ 
& & \hspace{-3mm} \int^{r_2}_{0} \hspace{-1mm} d r_3 \hspace{0.5mm} W(r_3, r_2)
\big[\opa{a} \opc{b}{\Psi}_{W}(r_1) \opa{b}\opc{c}{\Psi}_{W}(r_2)\opa{c}{\Psi}_{W}(r_3) \big] 
\opc{a} \Theta(\hat{\mathbf{n}}) \hspace{-0.5mm} \Bigg\}, \\ \nonumber
B&=& \hspace{-1mm} - \Bigg\{ 4  \int^{r_{s}}_{0} \hspace{-1mm} d r_1 \hspace{0.5mm} W(r_1, r_{s})  
\hspace{-1mm} \int^{r_1}_{0} \hspace{-1mm} d r_2 \hspace{0.5mm} W(r_2, r_1)  \big[\opa{a}\opc{b}\opc{c}{\Psi}_{W}(r_1) \opa{b}{\Psi}_{W}(r_2) \big] \\ 
& & \int^{r_1}_{0} \hspace{-1mm} d r_3 \hspace{1mm} W(r_3, r_1)
\opa{c}{\Psi}_{W}(r_3) 
\opc{a} \Theta(\hat{\mathbf{n}}) \Bigg\}, \\ \nonumber
C &=& - \Bigg\{ 8  \int^{r_{s}}_{0} \hspace{-1mm} d r_1 \hspace{1mm} W(r_1, r_{s})  \opa{a}{\Psi}_{W}(r_1) 
\hspace{-1mm} \int^{r_1}_{0} \hspace{-1mm} d r_2 \hspace{1mm} W(r_2, r_{s})  \\ 
& & \int^{r_2}_{0} \hspace{-1mm} d r_3 \hspace{1mm} W(r_3, r_2) 
 \opa{b}\opc{c}{\Psi}_{W}(r_2)\opa{c}{\Psi}_{W}(r_3) 
\opc{a} \opc{b} \Theta(\hat{\mathbf{n}}) \Bigg\}, \\ \nonumber
D&=& - \Bigg\{ 8  \int^{r_{s}}_{0} \hspace{-1mm} d r_1 \hspace{1mm} W(r_1, r_{s})   
\hspace{-1mm} \int^{r_1}_{0} \hspace{-1mm} d r_2 \hspace{1mm} W(r_2, r_1)  \opa{a}\opc{c}{\Psi}_{W}(r_1) \opa{c}{\Psi}_{W}(r_2) \\ 
& &   \int^{r_2}_{0} \hspace{-1mm} d r_3 \hspace{1mm} W(r_3, r_{s})  
\opa{b}{\Psi}_{W}(r_3) 
\opc{a} \opc{b} \Theta(\hat{\mathbf{n}}) \Bigg\}, \\ \nonumber
E&=& - \Bigg\{ 8  \int^{r_{s}}_{0} \hspace{-1mm} d r_1 \hspace{1mm} W(r_1, r_{s})   
\hspace{-1mm} \int^{r_1}_{0} \hspace{-1mm} d r_2 \hspace{1mm} W(r_2, r_{s})   \\ 
& &   \int^{r_2}_{0} \hspace{-1mm} d r_3 \hspace{1mm} W(r_3, r_1)  
\opa{a}\opc{c}{\Psi}_{W}(r_1) \opa{b}{\Psi}_{W}(r_2) \opa{c}{\Psi}_{W}(r_3) 
\opc{a} \opc{b} \Theta(\hat{\mathbf{n}}) \Bigg\} , \\ \nonumber
F&=& - \Bigg\{ {4\over 3}  \bigg[ \int^{r_{s}}_{0} \hspace{-2mm} d r_1 \hspace{1mm} W(r_1, r_{s}) \opa{a}{\Psi}_{W}(r_1) \bigg]
 \hspace{-1mm} \bigg[ \int^{r_{s}}_{0} \hspace{-2mm} d r_2 \hspace{1mm} W(r_2, r_{s})  \opa{b}{\Psi}_{W}(r_2) \bigg] \\
& & \bigg[ \int^{r_{s}}_{0} \hspace{-2mm} d r_3 \hspace{1mm} W(r_3, r_{s})  \opa{c}{\Psi}_{W}(r_3) \bigg]
\opa{a} \opc{b} \opc{c} \Theta(\hat{\mathbf{n}}) \Bigg\} . 
\end{eqnarray}

By comparing these solutions term by term to the remapping solution, \eqn{Pratten fourth order}, one can see that three of the terms coincide; \(A = A'\), \(B = B'\) and \(F = F'\). However, as previously noted, terms \(C'\) and \(D'\) are symmetric in the remapping solution, and are represented by the diagram 
\begin{tikzpicture}[baseline=(current bounding box.west)]
\fill (0,0) circle[radius=\dotsize]; \draw (0,0) node[below]{$r_s$};
\fill (0.6,0) \Square{\dotsize}; \draw (0.6,0) node[below]{$r_1$};
\fill (1.2,0) \Square{\dotsize}; \draw (1.2,0) node[below]{$r_2$};
\fill (1.8,0) \Square{\dotsize}; \draw (1.8,0) node[below]{$r_3$};
\draw[dashed] (0,0) to [out=60, in=120] (0.6,0);
\draw[dashed] (0.0,0) to [out=60, in=120] (1.2,0);
\draw[dashed] (1.2,0) to [out=60, in=120] (1.8,0);
\end{tikzpicture} 
and its symmetric counterpart, whereas in the Boltzmann solution terms \(C\) and \(D\) differ due to the time ordering information encoded in the nested integrals. As \(C'\) and \(D'\) are symmetric they can be added together to form a single term:
\begin{eqnarray}
C' + D'  \equiv G = &&- \Bigg\{ 8  \int^{r_{s}}_{0} \hspace{-1mm} d r_1 \hspace{1mm} W(r_1, r_{s})  \opa{a}\opc{c}{\Psi}_{W}(r_1)
\hspace{-1mm} \int^{r_1}_{0} \hspace{-1mm} d r_2 \hspace{1mm} W(r_2, r_1)  \opa{c}{\Psi}_{W}(r_2) 
\nonumber \\
& &  \int^{r_{s}}_{0} \hspace{-1mm} d r_3 \hspace{1mm} W(r_3, r_{s}) 
\opa{b}{\Psi}_{W}(r_3) 
 \Bigg\} ,
\end{eqnarray}
where we have simplified the expression by dropping the derivatives of the first order anisotropies (we are working with the expression for the deflection angle product: \({\alpha}^{a [I]} {\alpha}^{b [II]}\)).

To prove the equivalence of the Boltzmann equation and remapping methods we must show that this term \(G\) is equivalent to the sum of the terms \(C\), \(D\) and \(E\), which written in this simplified form are
\begin{eqnarray}
C &=& - \Bigg\{ 8  \int^{r_{s}}_{0} \hspace{-1mm} d r_1 \hspace{1mm} W(r_1, r_{s})  \opa{a}{\Psi}_{W}(r_1) 
\hspace{-1mm} \int^{r_1}_{0} \hspace{-1mm} d r_2 \hspace{1mm} W(r_2, r_{s})  \opa{b}\opc{c}{\Psi}_{W}(r_2) \nonumber \\ 
& & \int^{r_2}_{0} \hspace{-1mm} d r_3 \hspace{1mm} W(r_3, r_2)
\opa{c}{\Psi}_{W}(r_3) 
 \Bigg\}, \\ \nonumber
D&=& - \Bigg\{ 8  \int^{r_{s}}_{0} \hspace{-1mm} d r_1 \hspace{1mm} W(r_1, r_{s})  \opa{a}\opc{c}{\Psi}_{W}(r_1)  
\hspace{-1mm} \int^{r_1}_{0} \hspace{-1mm} d r_2 \hspace{1mm} W(r_2, r_1)  \opa{c}{\Psi}_{W}(r_2) \\ 
& &   \int^{r_2}_{0} \hspace{-1mm} d r_3 \hspace{1mm} W(r_3, r_{s}) 
\opa{b}{\Psi}_{W}(r_3) 
 \Bigg\}, \\ \nonumber
E&=& - \Bigg\{ 8  \int^{r_{s}}_{0} \hspace{-1mm} d r_1 \hspace{1mm} W(r_1, r_{s}) \opa{a}\opc{c}{\Psi}_{W}(r_1)
\hspace{-1mm} \int^{r_1}_{0} \hspace{-1mm} d r_2 \hspace{1mm} W(r_2, r_{s}) \opa{b}{\Psi}_{W}(r_2)  \\ 
& &   \int^{r_2}_{0} \hspace{-1mm} d r_3 \hspace{1mm} W(r_3, r_1)  
  \opa{c}{\Psi}_{W}(r_3) 
 \Bigg\} . 
\end{eqnarray}
By examination we can see that term \(D\) is most similar in form to \(G\) and we can in fact immediately recover it by splitting the third integral in \(G\):
\begin{eqnarray}
G &=& - \Bigg\{ 8  \int^{r_{s}}_{0} \hspace{-1mm} d r_1 \hspace{1mm} W(r_1, r_{s})  \opa{a}\opc{c}{\Psi}_{W}(r_1)
\hspace{-1mm} \int^{r_1}_{0} \hspace{-1mm} d r_2 \hspace{1mm} W(r_2, r_1)  \opa{c}{\Psi}_{W}(r_2) 
\nonumber \\
& &  \bigg[ \underbrace{\int^{r_{2}}_{0} \hspace{-1mm} d r_3 \hspace{1mm} W(r_3, r_{s}) \opa{b}{\Psi}_{W}(r_3) }_{1}
+ \underbrace{\int^{r_{s}}_{r_{2}} \hspace{-1mm} d r_3 \hspace{1mm} W(r_3, r_{s}) \opa{b}{\Psi}_{W}(r_3) }_{2} \bigg]
 \Bigg\} ,
\end{eqnarray}
where we have split \(G\) in to two terms, \(G1\) and \(G2\):
\begin{eqnarray}
G1 &=& - \Bigg\{ 8  \int^{r_{s}}_{0} \hspace{-1mm} d r_1 \hspace{1mm} W(r_1, r_{s})  \opa{a}\opc{c}{\Psi}_{W}(r_1)  
\hspace{-1mm} \int^{r_1}_{0} \hspace{-1mm} d r_2 \hspace{1mm} W(r_2, r_1)  \opa{c}{\Psi}_{W}(r_2) \nonumber \\ 
& &   \int^{r_2}_{0} \hspace{-1mm} d r_3 \hspace{1mm} W(r_3, r_{s}) 
\opa{b}{\Psi}_{W}(r_3) 
 \Bigg\},~\mathrm{and}~ \\ \nonumber
G2 &=& - \Bigg\{ 8  \int^{r_{s}}_{0} \hspace{-1mm} d r_1 \hspace{1mm} W(r_1, r_{s})  \opa{a}\opc{c}{\Psi}_{W}(r_1)  
\hspace{-1mm} \int^{r_1}_{0} \hspace{-1mm} d r_2 \hspace{1mm} W(r_2, r_1)  \opa{c}{\Psi}_{W}(r_2) \nonumber \\ 
& &   \int^{r_{s}}_{r_{2}} \hspace{-1mm} d r_3 \hspace{1mm} W(r_3, r_{s}) 
\opa{b}{\Psi}_{W}(r_3) 
 \Bigg\} .
\end{eqnarray}
We see that \(G1 = D\). Focusing on \(G2\), we can move closer to term \(E\) by first exchanging the labels of \(r_2\) and \(r_3\) to obtain the same form of the lensing efficiency functions, \(W(r_1, r_2)\)
\begin{eqnarray}
G2 &=& - \Bigg\{ 8  \int^{r_{s}}_{0} \hspace{-1mm} d r_1 \hspace{1mm} W(r_1, r_{s})  \opa{a}\opc{c}{\Psi}_{W}(r_1)  
\hspace{-1mm} \int^{r_1}_{0} \hspace{-1mm} d r_3 \hspace{1mm} W(r_3, r_1)  \opa{c}{\Psi}_{W}(r_3) \nonumber \\ 
& &   \int^{r_{s}}_{r_{3}} \hspace{-1mm} d r_2 \hspace{1mm} W(r_2, r_{s}) 
\opa{b}{\Psi}_{W}(r_2) 
 \Bigg\} .
\end{eqnarray}
If we then split the middle integral
\begin{eqnarray}
G2 &=& - \Bigg\{ 8  \int^{r_{s}}_{0} \hspace{-1mm} d r_1 \hspace{1mm} W(r_1, r_{s})  \opa{a}\opc{c}{\Psi}_{W}(r_1)  
\hspace{-1mm} \bigg[ \underbrace{\int^{r_s}_{0} \hspace{-1mm} d r_3 \hspace{1mm} W(r_3, r_1)  \opa{c}{\Psi}_{W}(r_3) }_{a} \nonumber \\ 
& &  - \underbrace{\int^{r_s}_{r_1} \hspace{-1mm} d r_3 \hspace{1mm} W(r_3, r_1)  \opa{c}{\Psi}_{W}(r_3) }_{b} \bigg]
\int^{r_{s}}_{r_{3}} \hspace{-1mm} d r_2 \hspace{1mm} W(r_2, r_{s}) 
\opa{b}{\Psi}_{W}(r_2) 
 \Bigg\} , 
\end{eqnarray}
we obtain two terms, \(G2a\) and \(G2b\):
\begin{eqnarray}
G2a &=& - \Bigg\{ 8  \int^{r_{s}}_{0} \hspace{-1mm} d r_1 \hspace{1mm} W(r_1, r_{s})  \opa{a}\opc{c}{\Psi}_{W}(r_1)  
\hspace{-1mm} \int^{r_s}_{0} \hspace{-1mm} d r_3 \hspace{1mm} W(r_3, r_1)  \opa{c}{\Psi}_{W}(r_3)  \nonumber \\ 
& & \int^{r_{s}}_{r_{3}} \hspace{-1mm} d r_2 \hspace{1mm} W(r_2, r_{s}) 
\opa{b}{\Psi}_{W}(r_2) 
 \Bigg\} , ~\mathrm{and}~ \\ 
G2b &=&  \Bigg\{ 8  \int^{r_{s}}_{0} \hspace{-1mm} d r_1 \hspace{1mm} W(r_1, r_{s})  \opa{a}\opc{c}{\Psi}_{W}(r_1)  
\hspace{-1mm} \int^{r_s}_{r_1} \hspace{-1mm} d r_3 \hspace{1mm} W(r_3, r_1)  \opa{c}{\Psi}_{W}(r_3) \nonumber \\ 
& &  \int^{r_{s}}_{r_{3}} \hspace{-1mm} d r_2 \hspace{1mm} W(r_2, r_{s}) 
\opa{b}{\Psi}_{W}(r_2) 
 \Bigg\} .
\end{eqnarray}
We can then rearrange the order of integration of \(G2a\) by using \eqn{rearrange order} 
\begin{eqnarray}
\int^{r_{M+1}}_{0} d r_{M} \int^{r_{M}}_{0} d \tilde{r} = \int^{r_{M+1}}_{0} d \tilde{r} \int^{r_{M+1}}_{\tilde{r}} d r_{M}
\end{eqnarray}
to obtain
\begin{eqnarray}
G2a &=& - \Bigg\{ 8  \int^{r_{s}}_{0} \hspace{-1mm} d r_1 \hspace{1mm} W(r_1, r_{s})  \opa{a}\opc{c}{\Psi}_{W}(r_1)  
\hspace{-1mm} \int^{r_{s}}_{0} \hspace{-1mm} d r_2 \hspace{1mm} W(r_2, r_{s}) \opa{b}{\Psi}_{W}(r_2)   \nonumber \\ 
& &  \int^{r_2}_{0} \hspace{-1mm} d r_3 \hspace{1mm} W(r_3, r_1)  \opa{c}{\Psi}_{W}(r_3)
 \Bigg\} .  
\end{eqnarray}
The middle integral can then be split once more to recover term \(E\):
\begin{eqnarray}
G2a &=& - \Bigg\{ 8  \int^{r_{s}}_{0} \hspace{-1mm} d r_1 \hspace{1mm} W(r_1, r_{s})  \opa{a}\opc{c}{\Psi}_{W}(r_1)  
\hspace{-1mm} \bigg[ \underbrace{\int^{r_{1}}_{0} \hspace{-1mm} d r_2 \hspace{1mm} W(r_2, r_{s}) \opa{b}{\Psi}_{W}(r_2) }_{1}   \nonumber \\ 
& & + \underbrace{ \int^{r_{s}}_{r_{1}} \hspace{-1mm} d r_2 \hspace{1mm} W(r_2, r_{s}) \opa{b}{\Psi}_{W}(r_2) \bigg] }_{2}
\int^{r_2}_{0} \hspace{-1mm} d r_3 \hspace{1mm} W(r_3, r_1)  \opa{c}{\Psi}_{W}(r_3)
 \Bigg\} , 
\end{eqnarray}
giving us a pair of terms
\begin{eqnarray}
G2a1 &=& - \Bigg\{ 8  \int^{r_{s}}_{0} \hspace{-1mm} d r_1 \hspace{1mm} W(r_1, r_{s})  \opa{a}\opc{c}{\Psi}_{W}(r_1)  
\hspace{-1mm} \int^{r_{1}}_{0} \hspace{-1mm} d r_2 \hspace{1mm} W(r_2, r_{s}) \opa{b}{\Psi}_{W}(r_2)    \nonumber \\ 
& & \int^{r_2}_{0} \hspace{-1mm} d r_3 \hspace{1mm} W(r_3, r_1)  \opa{c}{\Psi}_{W}(r_3)
 \Bigg\} , ~\mathrm{and}~\\
G2a2 &=& - \Bigg\{ 8  \int^{r_{s}}_{0} \hspace{-1mm} d r_1 \hspace{1mm} W(r_1, r_{s})  \opa{a}\opc{c}{\Psi}_{W}(r_1)  
\hspace{-1mm}  \int^{r_{s}}_{r_{1}} \hspace{-1mm} d r_2 \hspace{1mm} W(r_2, r_{s}) \opa{b}{\Psi}_{W}(r_2)  \nonumber \\ 
& & \int^{r_2}_{0} \hspace{-1mm} d r_3 \hspace{1mm} W(r_3, r_1)  \opa{c}{\Psi}_{W}(r_3)
 \Bigg\} , 
\end{eqnarray}
where it can be seen that \(G2a1 = E\). We now have one term left from the Boltzmann equation solution to recover, term \(C\), and two terms from the remapping method from which to do so, \(G2b\) and \(G2a2\). \(G2a2\) can be brought closer to the form of \(E\) by rearranging the order of integration using \eqn{rearrange order} so that
\begin{eqnarray}
G2a2 &=& - \Bigg\{ 8 \int^{r_{s}}_{0} \hspace{-1mm} d r_2 \hspace{1mm} W(r_2, r_{s}) \opa{b}{\Psi}_{W}(r_2)   
\hspace{-1mm}  \int^{r_{2}}_{0} \hspace{-1mm} d r_1 \hspace{1mm} W(r_1, r_{s})  \opa{a}\opc{c}{\Psi}_{W}(r_1) \nonumber \\ 
& & \int^{r_2}_{0} \hspace{-1mm} d r_3 \hspace{1mm} W(r_3, r_1)  \opa{c}{\Psi}_{W}(r_3)
 \Bigg\} .
\end{eqnarray}
Exchanging the labels of \(r_1\) and \(r_2\) then gives the correct form of the lensing effiency function
\begin{eqnarray}
G2a2 &=& - \Bigg\{ 8 \int^{r_{s}}_{0} \hspace{-1mm} d r_1 \hspace{1mm} W(r_1, r_{s}) \opa{b}{\Psi}_{W}(r_1)   
\hspace{-1mm}  \int^{r_{1}}_{0} \hspace{-1mm} d r_2 \hspace{1mm} W(r_2, r_{s})  \opa{a}\opc{c}{\Psi}_{W}(r_2) \nonumber \\ 
& & \int^{r_1}_{0} \hspace{-1mm} d r_3 \hspace{1mm} W(r_3, r_2)  \opa{c}{\Psi}_{W}(r_3)
 \Bigg\} .
\end{eqnarray}
Finally, the we split the third integral
\begin{eqnarray}
G2a2 &=& - \Bigg\{ 8 \int^{r_{s}}_{0} \hspace{-1mm} d r_1 \hspace{1mm} W(r_1, r_{s}) \opa{b}{\Psi}_{W}(r_1)   
\hspace{-1mm}  \int^{r_{1}}_{0} \hspace{-1mm} d r_2 \hspace{1mm} W(r_2, r_{s})  \opa{a}\opc{c}{\Psi}_{W}(r_2) \nonumber \\ 
& & \bigg[ \underbrace{ \int^{r_2}_{0} \hspace{-1mm} d r_3 \hspace{1mm} W(r_3, r_2)  \opa{c}{\Psi}_{W}(r_3) }_{a}
- \underbrace{ \int^{r_2}_{r_1} \hspace{-1mm} d r_3 \hspace{1mm} W(r_3, r_2)  \opa{c}{\Psi}_{W}(r_3) }_{b} \bigg]
 \Bigg\} ,
\end{eqnarray}
to obtain two terms
\begin{eqnarray}
G2a2a &=& - \Bigg\{ 8 \int^{r_{s}}_{0} \hspace{-1mm} d r_1 \hspace{1mm} W(r_1, r_{s}) \opa{b}{\Psi}_{W}(r_1)   
\hspace{-1mm}  \int^{r_{1}}_{0} \hspace{-1mm} d r_2 \hspace{1mm} W(r_2, r_{s})  \opa{a}\opc{c}{\Psi}_{W}(r_2) \nonumber \\ 
& &  \int^{r_2}_{0} \hspace{-1mm} d r_3 \hspace{1mm} W(r_3, r_2)  \opa{c}{\Psi}_{W}(r_3) 
 \Bigg\} ,~\mathrm{and} \\
G2a2b &=& \Bigg\{ 8 \int^{r_{s}}_{0} \hspace{-1mm} d r_1 \hspace{1mm} W(r_1, r_{s}) \opa{b}{\Psi}_{W}(r_1)   
\hspace{-1mm}  \int^{r_{1}}_{0} \hspace{-1mm} d r_2 \hspace{1mm} W(r_2, r_{s})  \opa{a}\opc{c}{\Psi}_{W}(r_2) \nonumber \\ 
& & \int^{r_2}_{r_1} \hspace{-1mm} d r_3 \hspace{1mm} W(r_3, r_2)  \opa{c}{\Psi}_{W}(r_3) 
 \Bigg\} .
\end{eqnarray}
The first term is in fact the final recovered term, \(G2a2a = C\). We have recovered all of the necessary terms, but are left with two residual terms, \(G2b\) and \(G2a2b\) which must cancel to prove equivalence between our two results:
\begin{eqnarray}
G2b &=&  \Bigg\{ 8  \int^{r_{s}}_{0} \hspace{-1mm} d r_1 \hspace{1mm} W(r_1, r_{s})  \opa{a}\opc{c}{\Psi}_{W}(r_1)  
\hspace{-1mm} \int^{r_s}_{r_1} \hspace{-1mm} d r_3 \hspace{1mm} W(r_3, r_1)  \opa{c}{\Psi}_{W}(r_3) \nonumber \\ 
& &  \int^{r_{s}}_{r_{3}} \hspace{-1mm} d r_2 \hspace{1mm} W(r_2, r_{s}) \opa{b}{\Psi}_{W}(r_2) 
 \Bigg\} , ~\mathrm{and}\\
G2a2b &=& \Bigg\{ -8 \int^{r_{s}}_{0} \hspace{-1mm} d r_1 \hspace{1mm} W(r_1, r_{s}) \opa{b}{\Psi}_{W}(r_1)   
\hspace{-1mm}  \int^{r_{1}}_{0} \hspace{-1mm} d r_2 \hspace{1mm} W(r_2, r_{s})  \opa{a}\opc{c}{\Psi}_{W}(r_2) \nonumber \\ 
& & \int^{r_1}_{r_2} \hspace{-1mm} d r_3 \hspace{1mm} W(r_3, r_2)  \opa{c}{\Psi}_{W}(r_3) 
 \Bigg\} .
\end{eqnarray}
We begin bringing them to the same form by rearranging the order of integration of \(G2b\) using \eqn{rearrange order}
\begin{eqnarray}
G2b &=&  \Bigg\{ 8 \int^{r_s}_{0} \hspace{-1mm} d r_3 \hspace{1mm} W(r_3, r_1)  \opa{c}{\Psi}_{W}(r_3)
\hspace{-1mm} \int^{r_{3}}_{0} \hspace{-1mm} d r_1 \hspace{1mm} W(r_1, r_{s})  \opa{a}\opc{c}{\Psi}_{W}(r_1)   \nonumber \\ 
& &  \int^{r_{s}}_{r_{3}} \hspace{-1mm} d r_2 \hspace{1mm} W(r_2, r_{s}) \opa{b}{\Psi}_{W}(r_2) 
 \Bigg\} ,
\end{eqnarray}
then repeating this operation to bring the third integral out to the front:
\begin{eqnarray}
G2b &=&  \Bigg\{ 8 \int^{r_{s}}_{0} \hspace{-1mm} d r_2 \hspace{1mm} W(r_2, r_{s}) \opa{b}{\Psi}_{W}(r_2) 
\hspace{-1mm} \int^{r_2}_{0} \hspace{-1mm} d r_3 \hspace{1mm} W(r_3, r_1)  \opa{c}{\Psi}_{W}(r_3) \nonumber \\ 
& & \int^{r_{3}}_{0} \hspace{-1mm} d r_1 \hspace{1mm} W(r_1, r_{s})  \opa{a}\opc{c}{\Psi}_{W}(r_1) 
 \Bigg\} .
\end{eqnarray}
Turning to term \(G2a2b\), we exchange the labels \(r_1\), \(r_2\):
\begin{eqnarray}
G2a2b &=& \Bigg\{ -8 \int^{r_{s}}_{0} \hspace{-1mm} d r_2 \hspace{1mm} W(r_2, r_{s}) \opa{b}{\Psi}_{W}(r_2)   
\hspace{-1mm}  \int^{r_{2}}_{0} \hspace{-1mm} d r_1 \hspace{1mm} W(r_1, r_{s})  \opa{a}\opc{c}{\Psi}_{W}(r_1) \nonumber \\ 
& & \int^{r_2}_{r_1} \hspace{-1mm} d r_3 \hspace{1mm} W(r_3, r_1)  \opa{c}{\Psi}_{W}(r_3) 
 \Bigg\} ,
\end{eqnarray}
then rearrange the order of integration to obtain
\begin{eqnarray}
G2a2b &=& \Bigg\{ -8 \int^{r_{s}}_{0} \hspace{-1mm} d r_2 \hspace{1mm} W(r_2, r_{s}) \opa{b}{\Psi}_{W}(r_2)   
\hspace{-1mm} \int^{r_2}_{0} \hspace{-1mm} d r_3 \hspace{1mm} W(r_3, r_1)  \opa{c}{\Psi}_{W}(r_3)  \nonumber \\ 
& & \int^{r_{3}}_{0} \hspace{-1mm} d r_1 \hspace{1mm} W(r_1, r_{s})  \opa{a}\opc{c}{\Psi}_{W}(r_1)
 \Bigg\} .
\end{eqnarray}
It can then be seen that \(G2b + G2a2b = 0\), proving that the Boltzmann equation and the remapping method are equivalent to fourth order.

\section{Discussion and Conclusion} \label{sect:conclusion}

In this paper, we prove the equivalence of the canonical weak lensing remapping ansatz and the Boltzmann approach, up to 4th order. We believe that this equivalence is general to all orders, although a general proof eludes us at this moment.  The use of the Boltzmann approach allows us to shed insight into the so-called ``post-Born'' effects of weak lensing, allowing us to demonstrate that the such effects are simply lens-lens coupling, and the Born Approximation is still valid -- all the line integrals are performed along the unperturbed photon path $\bar{\bx}$. The Boltzmann approach also identifies the exact point where true post-Born effects could be incorporated -- this lies in the 2nd term of the Liouville expansion \eqn{eqn:Liouville_intensity} which is presently neglected. As argued in \cite{Su:2014mga} we expect this term to be non-dominant.

Ultimately, whether or not to call higher order weak lensing effects ``post-Born'' is a question of nomenclature. As discussed in the literature \cite{Beck:2018wud,Pratten:2016dsm,Marozzi:2016und,Marozzi:2016uob,Fabbian:2017wfp,Takahashi:2017hjr,Fabbian:2019tik}, 4th order lensing effects on the CMB is very small. A full treatment \emph{ala} Boltzmann has shown that this effect is $\sim0.01\%$ of the CMB at the most \cite{Su:2014mga}, thus is negligible except perhaps as a systematic. On the other hand, in studies of weak lensing of the galaxy power at higher orders follow very similar ``remapping'' approaches \cite{Krause:2009yr} -- instead of the unlensed CMB as the source, we replace the $\Theta^{[I]}$ power spectrum with the equivalent unlensed galaxy power spectrum. Up to 4th order, the effects are estimated at $1-5\sigma$ level \cite{Krause:2009yr,Petri:2016qya} and hence a full accounting of all the lensing terms is potentially important in future studies \cite{Bohm:2019bek}. 

\acknowledgments

We would like to thank  Tasos Avgoustidis, Alex Jenkins, David Mulryne, and Simon Su for their help and comments on this work. EAL was supported an STFC AGP grant no. ST/L000717/1 for part of this work.

\bibliography{lensingref2.bib}

\begin{thebibliography}{56}%
\makeatletter
\providecommand \@ifxundefined [1]{%
 \@ifx{#1\undefined}
}%
\providecommand \@ifnum [1]{%
 \ifnum #1\expandafter \@firstoftwo
 \else \expandafter \@secondoftwo
 \fi
}%
\providecommand \@ifx [1]{%
 \ifx #1\expandafter \@firstoftwo
 \else \expandafter \@secondoftwo
 \fi
}%
\providecommand \natexlab [1]{#1}%
\providecommand \enquote  [1]{``#1''}%
\providecommand \bibnamefont  [1]{#1}%
\providecommand \bibfnamefont [1]{#1}%
\providecommand \citenamefont [1]{#1}%
\providecommand \href@noop [0]{\@secondoftwo}%
\providecommand \href [0]{\begingroup \@sanitize@url \@href}%
\providecommand \@href[1]{\@@startlink{#1}\@@href}%
\providecommand \@@href[1]{\endgroup#1\@@endlink}%
\providecommand \@sanitize@url [0]{\catcode `\\12\catcode `\$12\catcode
  `\&12\catcode `\#12\catcode `\^12\catcode `\_12\catcode `\%12\relax}%
\providecommand \@@startlink[1]{}%
\providecommand \@@endlink[0]{}%
\providecommand \url  [0]{\begingroup\@sanitize@url \@url }%
\providecommand \@url [1]{\endgroup\@href {#1}{\urlprefix }}%
\providecommand \urlprefix  [0]{URL }%
\providecommand \Eprint [0]{\href }%
\providecommand \doibase [0]{http://dx.doi.org/}%
\providecommand \selectlanguage [0]{\@gobble}%
\providecommand \bibinfo  [0]{\@secondoftwo}%
\providecommand \bibfield  [0]{\@secondoftwo}%
\providecommand \translation [1]{[#1]}%
\providecommand \BibitemOpen [0]{}%
\providecommand \bibitemStop [0]{}%
\providecommand \bibitemNoStop [0]{.\EOS\space}%
\providecommand \EOS [0]{\spacefactor3000\relax}%
\providecommand \BibitemShut  [1]{\csname bibitem#1\endcsname}%
\let\auto@bib@innerbib\@empty
\bibitem [{\citenamefont {Su}\ and\ \citenamefont {Lim}(2014)}]{Su:2014mga}%
  \BibitemOpen
  \bibfield  {author} {\bibinfo {author} {\bibfnamefont {S.~C.}\ \bibnamefont
  {Su}}\ and\ \bibinfo {author} {\bibfnamefont {E.~A.}\ \bibnamefont {Lim}},\
  }\href {\doibase 10.1103/PhysRevD.89.123006} {\bibfield  {journal} {\bibinfo
  {journal} {Phys. Rev.}\ }\textbf {\bibinfo {volume} {D89}},\ \bibinfo {pages}
  {123006} (\bibinfo {year} {2014})},\ \Eprint {http://arxiv.org/abs/1401.5737}
  {arXiv:1401.5737 [astro-ph.CO]} \BibitemShut {NoStop}%
\bibitem [{\citenamefont {Pyne}\ and\ \citenamefont
  {Carroll}(1996)}]{Pyne:1995bs}%
  \BibitemOpen
  \bibfield  {author} {\bibinfo {author} {\bibfnamefont {T.}~\bibnamefont
  {Pyne}}\ and\ \bibinfo {author} {\bibfnamefont {S.~M.}\ \bibnamefont
  {Carroll}},\ }\href {\doibase 10.1103/PhysRevD.53.2920} {\bibfield  {journal}
  {\bibinfo  {journal} {Phys. Rev. D}\ }\textbf {\bibinfo {volume} {53}},\
  \bibinfo {pages} {2920} (\bibinfo {year} {1996})},\ \Eprint
  {http://arxiv.org/abs/astro-ph/9510041} {arXiv:astro-ph/9510041} \BibitemShut
  {NoStop}%
\bibitem [{\citenamefont {Kaiser}(1998)}]{Kaiser:1996tp}%
  \BibitemOpen
  \bibfield  {author} {\bibinfo {author} {\bibfnamefont {N.}~\bibnamefont
  {Kaiser}},\ }\href {\doibase 10.1086/305515} {\bibfield  {journal} {\bibinfo
  {journal} {Astrophys. J.}\ }\textbf {\bibinfo {volume} {498}},\ \bibinfo
  {pages} {26} (\bibinfo {year} {1998})},\ \Eprint
  {http://arxiv.org/abs/astro-ph/9610120} {arXiv:astro-ph/9610120 [astro-ph]}
  \BibitemShut {NoStop}%
\bibitem [{\citenamefont {Wu}\ \emph {et~al.}(2019)\citenamefont {Wu} \emph
  {et~al.}}]{Wu:2019hek}%
  \BibitemOpen
  \bibfield  {author} {\bibinfo {author} {\bibfnamefont {W.~L.~K.}\
  \bibnamefont {Wu}} \emph {et~al.},\ }\href {\doibase
  10.3847/1538-4357/ab4186} {\bibfield  {journal} {\bibinfo  {journal}
  {Astrophys. J.}\ }\textbf {\bibinfo {volume} {884}},\ \bibinfo {pages} {70}
  (\bibinfo {year} {2019})},\ \Eprint {http://arxiv.org/abs/1905.05777}
  {arXiv:1905.05777 [astro-ph.CO]} \BibitemShut {NoStop}%
\bibitem [{\citenamefont {Namikawa}\ \emph {et~al.}(2019)\citenamefont
  {Namikawa} \emph {et~al.}}]{Namikawa:2019gtc}%
  \BibitemOpen
  \bibfield  {author} {\bibinfo {author} {\bibfnamefont {T.}~\bibnamefont
  {Namikawa}} \emph {et~al.} (\bibinfo {collaboration} {POLARBEAR, HSC}),\
  }\href {\doibase 10.3847/1538-4357/ab3424} {\bibfield  {journal} {\bibinfo
  {journal} {Astrophys. J.}\ }\textbf {\bibinfo {volume} {882}},\ \bibinfo
  {pages} {62} (\bibinfo {year} {2019})},\ \Eprint
  {http://arxiv.org/abs/1904.02116} {arXiv:1904.02116 [astro-ph.CO]}
  \BibitemShut {NoStop}%
\bibitem [{\citenamefont {Aghanim}\ \emph {et~al.}(2020)\citenamefont {Aghanim}
  \emph {et~al.}}]{Aghanim:2018oex}%
  \BibitemOpen
  \bibfield  {author} {\bibinfo {author} {\bibfnamefont {N.}~\bibnamefont
  {Aghanim}} \emph {et~al.} (\bibinfo {collaboration} {Planck}),\ }\href
  {\doibase 10.1051/0004-6361/201833886} {\bibfield  {journal} {\bibinfo
  {journal} {Astron. Astrophys.}\ }\textbf {\bibinfo {volume} {641}},\ \bibinfo
  {pages} {A8} (\bibinfo {year} {2020})},\ \Eprint
  {http://arxiv.org/abs/1807.06210} {arXiv:1807.06210 [astro-ph.CO]}
  \BibitemShut {NoStop}%
\bibitem [{\citenamefont {{Blanchard}}\ and\ \citenamefont
  {{Schneider}}(1987)}]{1987A&A...184....1B}%
  \BibitemOpen
  \bibfield  {author} {\bibinfo {author} {\bibfnamefont {A.}~\bibnamefont
  {{Blanchard}}}\ and\ \bibinfo {author} {\bibfnamefont {J.}~\bibnamefont
  {{Schneider}}},\ }\href@noop {} {\bibfield  {journal} {\bibinfo  {journal}
  {Astronomy and Astrophysics}\ }\textbf {\bibinfo {volume} {184}},\ \bibinfo
  {pages} {1} (\bibinfo {year} {1987})}\BibitemShut {NoStop}%
\bibitem [{\citenamefont {Zaldarriaga}\ \emph {et~al.}(1997)\citenamefont
  {Zaldarriaga}, \citenamefont {Spergel},\ and\ \citenamefont
  {Seljak}}]{Zaldarriaga_1997}%
  \BibitemOpen
  \bibfield  {author} {\bibinfo {author} {\bibfnamefont {M.}~\bibnamefont
  {Zaldarriaga}}, \bibinfo {author} {\bibfnamefont {D.~N.}\ \bibnamefont
  {Spergel}}, \ and\ \bibinfo {author} {\bibfnamefont {U.}~\bibnamefont
  {Seljak}},\ }\href {\doibase 10.1086/304692} {\bibfield  {journal} {\bibinfo
  {journal} {The Astrophysical Journal}\ }\textbf {\bibinfo {volume} {488}},\
  \bibinfo {pages} {1–13} (\bibinfo {year} {1997})}\BibitemShut {NoStop}%
\bibitem [{\citenamefont {Lewis}(2005)}]{Lewis_2005}%
  \BibitemOpen
  \bibfield  {author} {\bibinfo {author} {\bibfnamefont {A.}~\bibnamefont
  {Lewis}},\ }\href {\doibase 10.1103/physrevd.71.083008} {\bibfield  {journal}
  {\bibinfo  {journal} {Physical Review D}\ }\textbf {\bibinfo {volume} {71}}
  (\bibinfo {year} {2005}),\ 10.1103/physrevd.71.083008}\BibitemShut {NoStop}%
\bibitem [{\citenamefont {Kaplinghat}\ \emph {et~al.}(2003)\citenamefont
  {Kaplinghat}, \citenamefont {Knox},\ and\ \citenamefont
  {Song}}]{Kaplinghat_2003}%
  \BibitemOpen
  \bibfield  {author} {\bibinfo {author} {\bibfnamefont {M.}~\bibnamefont
  {Kaplinghat}}, \bibinfo {author} {\bibfnamefont {L.}~\bibnamefont {Knox}}, \
  and\ \bibinfo {author} {\bibfnamefont {Y.-S.}\ \bibnamefont {Song}},\ }\href
  {\doibase 10.1103/physrevlett.91.241301} {\bibfield  {journal} {\bibinfo
  {journal} {Physical Review Letters}\ }\textbf {\bibinfo {volume} {91}}
  (\bibinfo {year} {2003}),\ 10.1103/physrevlett.91.241301}\BibitemShut
  {NoStop}%
\bibitem [{\citenamefont {Liu}\ \emph {et~al.}(2016)\citenamefont {Liu},
  \citenamefont {Hill}, \citenamefont {Sherwin}, \citenamefont {Petri},
  \citenamefont {B\"ohm},\ and\ \citenamefont {Haiman}}]{Liu:2016nfs}%
  \BibitemOpen
  \bibfield  {author} {\bibinfo {author} {\bibfnamefont {J.}~\bibnamefont
  {Liu}}, \bibinfo {author} {\bibfnamefont {J.~C.}\ \bibnamefont {Hill}},
  \bibinfo {author} {\bibfnamefont {B.~D.}\ \bibnamefont {Sherwin}}, \bibinfo
  {author} {\bibfnamefont {A.}~\bibnamefont {Petri}}, \bibinfo {author}
  {\bibfnamefont {V.}~\bibnamefont {B\"ohm}}, \ and\ \bibinfo {author}
  {\bibfnamefont {Z.}~\bibnamefont {Haiman}},\ }\href {\doibase
  10.1103/PhysRevD.94.103501} {\bibfield  {journal} {\bibinfo  {journal} {Phys.
  Rev. D}\ }\textbf {\bibinfo {volume} {94}},\ \bibinfo {pages} {103501}
  (\bibinfo {year} {2016})},\ \Eprint {http://arxiv.org/abs/1608.03169}
  {arXiv:1608.03169 [astro-ph.CO]} \BibitemShut {NoStop}%
\bibitem [{\citenamefont {Green}\ \emph {et~al.}(2017)\citenamefont {Green},
  \citenamefont {Meyers},\ and\ \citenamefont {van Engelen}}]{Green:2016cjr}%
  \BibitemOpen
  \bibfield  {author} {\bibinfo {author} {\bibfnamefont {D.}~\bibnamefont
  {Green}}, \bibinfo {author} {\bibfnamefont {J.}~\bibnamefont {Meyers}}, \
  and\ \bibinfo {author} {\bibfnamefont {A.}~\bibnamefont {van Engelen}},\
  }\href {\doibase 10.1088/1475-7516/2017/12/005} {\bibfield  {journal}
  {\bibinfo  {journal} {JCAP}\ }\textbf {\bibinfo {volume} {12}},\ \bibinfo
  {pages} {005} (\bibinfo {year} {2017})},\ \Eprint
  {http://arxiv.org/abs/1609.08143} {arXiv:1609.08143 [astro-ph.CO]}
  \BibitemShut {NoStop}%
\bibitem [{\citenamefont {Bartelmann}\ and\ \citenamefont
  {Schneider}(2001)}]{Bartelmann:1999yn}%
  \BibitemOpen
  \bibfield  {author} {\bibinfo {author} {\bibfnamefont {M.}~\bibnamefont
  {Bartelmann}}\ and\ \bibinfo {author} {\bibfnamefont {P.}~\bibnamefont
  {Schneider}},\ }\href {\doibase 10.1016/S0370-1573(00)00082-X} {\bibfield
  {journal} {\bibinfo  {journal} {Phys. Rept.}\ }\textbf {\bibinfo {volume}
  {340}},\ \bibinfo {pages} {291} (\bibinfo {year} {2001})},\ \Eprint
  {http://arxiv.org/abs/astro-ph/9912508} {arXiv:astro-ph/9912508} \BibitemShut
  {NoStop}%
\bibitem [{\citenamefont {Hanson}\ \emph {et~al.}(2011)\citenamefont {Hanson},
  \citenamefont {Challinor}, \citenamefont {Efstathiou},\ and\ \citenamefont
  {Bielewicz}}]{Hanson_2011}%
  \BibitemOpen
  \bibfield  {author} {\bibinfo {author} {\bibfnamefont {D.}~\bibnamefont
  {Hanson}}, \bibinfo {author} {\bibfnamefont {A.}~\bibnamefont {Challinor}},
  \bibinfo {author} {\bibfnamefont {G.}~\bibnamefont {Efstathiou}}, \ and\
  \bibinfo {author} {\bibfnamefont {P.}~\bibnamefont {Bielewicz}},\ }\href
  {\doibase 10.1103/physrevd.83.043005} {\bibfield  {journal} {\bibinfo
  {journal} {Physical Review D}\ }\textbf {\bibinfo {volume} {83}} (\bibinfo
  {year} {2011}),\ 10.1103/physrevd.83.043005}\BibitemShut {NoStop}%
\bibitem [{\citenamefont {Lewis}\ \emph {et~al.}(2011)\citenamefont {Lewis},
  \citenamefont {Challinor},\ and\ \citenamefont {Hanson}}]{Lewis_2011}%
  \BibitemOpen
  \bibfield  {author} {\bibinfo {author} {\bibfnamefont {A.}~\bibnamefont
  {Lewis}}, \bibinfo {author} {\bibfnamefont {A.}~\bibnamefont {Challinor}}, \
  and\ \bibinfo {author} {\bibfnamefont {D.}~\bibnamefont {Hanson}},\ }\href
  {\doibase 10.1088/1475-7516/2011/03/018} {\bibfield  {journal} {\bibinfo
  {journal} {Journal of Cosmology and Astroparticle Physics}\ }\textbf
  {\bibinfo {volume} {2011}},\ \bibinfo {pages} {018–018} (\bibinfo {year}
  {2011})}\BibitemShut {NoStop}%
\bibitem [{\citenamefont {Zaldarriaga}\ and\ \citenamefont
  {Seljak}(1999)}]{Zaldarriaga:1998te}%
  \BibitemOpen
  \bibfield  {author} {\bibinfo {author} {\bibfnamefont {M.}~\bibnamefont
  {Zaldarriaga}}\ and\ \bibinfo {author} {\bibfnamefont {U.}~\bibnamefont
  {Seljak}},\ }\href {\doibase 10.1103/PhysRevD.59.123507} {\bibfield
  {journal} {\bibinfo  {journal} {Phys. Rev. D}\ }\textbf {\bibinfo {volume}
  {59}},\ \bibinfo {pages} {123507} (\bibinfo {year} {1999})},\ \Eprint
  {http://arxiv.org/abs/astro-ph/9810257} {arXiv:astro-ph/9810257} \BibitemShut
  {NoStop}%
\bibitem [{\citenamefont {Hu}(2001)}]{Hu:2001tn}%
  \BibitemOpen
  \bibfield  {author} {\bibinfo {author} {\bibfnamefont {W.}~\bibnamefont
  {Hu}},\ }\href {\doibase 10.1086/323253} {\bibfield  {journal} {\bibinfo
  {journal} {Astrophys. J. Lett.}\ }\textbf {\bibinfo {volume} {557}},\
  \bibinfo {pages} {L79} (\bibinfo {year} {2001})},\ \Eprint
  {http://arxiv.org/abs/astro-ph/0105424} {arXiv:astro-ph/0105424} \BibitemShut
  {NoStop}%
\bibitem [{\citenamefont {Hanson}\ \emph {et~al.}(2010)\citenamefont {Hanson},
  \citenamefont {Challinor},\ and\ \citenamefont {Lewis}}]{Hanson_2010}%
  \BibitemOpen
  \bibfield  {author} {\bibinfo {author} {\bibfnamefont {D.}~\bibnamefont
  {Hanson}}, \bibinfo {author} {\bibfnamefont {A.}~\bibnamefont {Challinor}}, \
  and\ \bibinfo {author} {\bibfnamefont {A.}~\bibnamefont {Lewis}},\ }\href
  {\doibase 10.1007/s10714-010-1036-y} {\bibfield  {journal} {\bibinfo
  {journal} {General Relativity and Gravitation}\ }\textbf {\bibinfo {volume}
  {42}},\ \bibinfo {pages} {2197–2218} (\bibinfo {year} {2010})}\BibitemShut
  {NoStop}%
\bibitem [{\citenamefont {Peloton}\ \emph {et~al.}(2017)\citenamefont
  {Peloton}, \citenamefont {Schmittfull}, \citenamefont {Lewis}, \citenamefont
  {Carron},\ and\ \citenamefont {Zahn}}]{Peloton:2016kbw}%
  \BibitemOpen
  \bibfield  {author} {\bibinfo {author} {\bibfnamefont {J.}~\bibnamefont
  {Peloton}}, \bibinfo {author} {\bibfnamefont {M.}~\bibnamefont
  {Schmittfull}}, \bibinfo {author} {\bibfnamefont {A.}~\bibnamefont {Lewis}},
  \bibinfo {author} {\bibfnamefont {J.}~\bibnamefont {Carron}}, \ and\ \bibinfo
  {author} {\bibfnamefont {O.}~\bibnamefont {Zahn}},\ }\href {\doibase
  10.1103/PhysRevD.95.043508} {\bibfield  {journal} {\bibinfo  {journal} {Phys.
  Rev. D}\ }\textbf {\bibinfo {volume} {95}},\ \bibinfo {pages} {043508}
  (\bibinfo {year} {2017})},\ \Eprint {http://arxiv.org/abs/1611.01446}
  {arXiv:1611.01446 [astro-ph.CO]} \BibitemShut {NoStop}%
\bibitem [{\citenamefont {Carron}\ and\ \citenamefont
  {Lewis}(2017)}]{Carron:2017mqf}%
  \BibitemOpen
  \bibfield  {author} {\bibinfo {author} {\bibfnamefont {J.}~\bibnamefont
  {Carron}}\ and\ \bibinfo {author} {\bibfnamefont {A.}~\bibnamefont {Lewis}},\
  }\href {\doibase 10.1103/PhysRevD.96.063510} {\bibfield  {journal} {\bibinfo
  {journal} {Phys. Rev. D}\ }\textbf {\bibinfo {volume} {96}},\ \bibinfo
  {pages} {063510} (\bibinfo {year} {2017})},\ \Eprint
  {http://arxiv.org/abs/1704.08230} {arXiv:1704.08230 [astro-ph.CO]}
  \BibitemShut {NoStop}%
\bibitem [{\citenamefont {Cooray}\ and\ \citenamefont
  {Hu}(2002)}]{Cooray:2002mj}%
  \BibitemOpen
  \bibfield  {author} {\bibinfo {author} {\bibfnamefont {A.}~\bibnamefont
  {Cooray}}\ and\ \bibinfo {author} {\bibfnamefont {W.}~\bibnamefont {Hu}},\
  }\href {\doibase 10.1086/340892} {\bibfield  {journal} {\bibinfo  {journal}
  {Astrophys. J.}\ }\textbf {\bibinfo {volume} {574}},\ \bibinfo {pages} {19}
  (\bibinfo {year} {2002})},\ \Eprint {http://arxiv.org/abs/astro-ph/0202411}
  {arXiv:astro-ph/0202411 [astro-ph]} \BibitemShut {NoStop}%
\bibitem [{\citenamefont {Song}\ \emph {et~al.}(2003)\citenamefont {Song},
  \citenamefont {Cooray}, \citenamefont {Knox},\ and\ \citenamefont
  {Zaldarriaga}}]{Song_2003}%
  \BibitemOpen
  \bibfield  {author} {\bibinfo {author} {\bibfnamefont {Y.}~\bibnamefont
  {Song}}, \bibinfo {author} {\bibfnamefont {A.}~\bibnamefont {Cooray}},
  \bibinfo {author} {\bibfnamefont {L.}~\bibnamefont {Knox}}, \ and\ \bibinfo
  {author} {\bibfnamefont {M.}~\bibnamefont {Zaldarriaga}},\ }\href {\doibase
  10.1086/375188} {\bibfield  {journal} {\bibinfo  {journal} {The Astrophysical
  Journal}\ }\textbf {\bibinfo {volume} {590}},\ \bibinfo {pages} {664–672}
  (\bibinfo {year} {2003})}\BibitemShut {NoStop}%
\bibitem [{\citenamefont {Hand}\ \emph {et~al.}(2015)\citenamefont {Hand},
  \citenamefont {Leauthaud}, \citenamefont {Das}, \citenamefont {Sherwin},
  \citenamefont {Addison}, \citenamefont {Bond}, \citenamefont {Calabrese},
  \citenamefont {Charbonnier}, \citenamefont {Devlin}, \citenamefont
  {Dunkley},\ and\ \citenamefont {et~al.}}]{Hand_2015}%
  \BibitemOpen
  \bibfield  {author} {\bibinfo {author} {\bibfnamefont {N.}~\bibnamefont
  {Hand}}, \bibinfo {author} {\bibfnamefont {A.}~\bibnamefont {Leauthaud}},
  \bibinfo {author} {\bibfnamefont {S.}~\bibnamefont {Das}}, \bibinfo {author}
  {\bibfnamefont {B.~D.}\ \bibnamefont {Sherwin}}, \bibinfo {author}
  {\bibfnamefont {G.~E.}\ \bibnamefont {Addison}}, \bibinfo {author}
  {\bibfnamefont {J.~R.}\ \bibnamefont {Bond}}, \bibinfo {author}
  {\bibfnamefont {E.}~\bibnamefont {Calabrese}}, \bibinfo {author}
  {\bibfnamefont {A.}~\bibnamefont {Charbonnier}}, \bibinfo {author}
  {\bibfnamefont {M.~J.}\ \bibnamefont {Devlin}}, \bibinfo {author}
  {\bibfnamefont {J.}~\bibnamefont {Dunkley}}, \ and\ \bibinfo {author}
  {\bibnamefont {et~al.}},\ }\href {\doibase 10.1103/physrevd.91.062001}
  {\bibfield  {journal} {\bibinfo  {journal} {Physical Review D}\ }\textbf
  {\bibinfo {volume} {91}} (\bibinfo {year} {2015}),\
  10.1103/physrevd.91.062001}\BibitemShut {NoStop}%
\bibitem [{\citenamefont {Singh}\ \emph {et~al.}(2016)\citenamefont {Singh},
  \citenamefont {Mandelbaum},\ and\ \citenamefont {Brownstein}}]{Singh_2016}%
  \BibitemOpen
  \bibfield  {author} {\bibinfo {author} {\bibfnamefont {S.}~\bibnamefont
  {Singh}}, \bibinfo {author} {\bibfnamefont {R.}~\bibnamefont {Mandelbaum}}, \
  and\ \bibinfo {author} {\bibfnamefont {J.~R.}\ \bibnamefont {Brownstein}},\
  }\href {\doibase 10.1093/mnras/stw2482} {\bibfield  {journal} {\bibinfo
  {journal} {Monthly Notices of the Royal Astronomical Society}\ }\textbf
  {\bibinfo {volume} {464}},\ \bibinfo {pages} {2120–2138} (\bibinfo {year}
  {2016})}\BibitemShut {NoStop}%
\bibitem [{\citenamefont {Harnois-Déraps}\ \emph {et~al.}(2017)\citenamefont
  {Harnois-Déraps}, \citenamefont {Tröster}, \citenamefont {Chisari},
  \citenamefont {Heymans}, \citenamefont {van Waerbeke}, \citenamefont
  {Asgari}, \citenamefont {Bilicki}, \citenamefont {Choi}, \citenamefont
  {Erben}, \citenamefont {Hildebrandt},\ and\ \citenamefont
  {et~al.}}]{Harnois_D_raps_2017}%
  \BibitemOpen
  \bibfield  {author} {\bibinfo {author} {\bibfnamefont {J.}~\bibnamefont
  {Harnois-Déraps}}, \bibinfo {author} {\bibfnamefont {T.}~\bibnamefont
  {Tröster}}, \bibinfo {author} {\bibfnamefont {N.~E.}\ \bibnamefont
  {Chisari}}, \bibinfo {author} {\bibfnamefont {C.}~\bibnamefont {Heymans}},
  \bibinfo {author} {\bibfnamefont {L.}~\bibnamefont {van Waerbeke}}, \bibinfo
  {author} {\bibfnamefont {M.}~\bibnamefont {Asgari}}, \bibinfo {author}
  {\bibfnamefont {M.}~\bibnamefont {Bilicki}}, \bibinfo {author} {\bibfnamefont
  {A.}~\bibnamefont {Choi}}, \bibinfo {author} {\bibfnamefont {T.}~\bibnamefont
  {Erben}}, \bibinfo {author} {\bibfnamefont {H.}~\bibnamefont {Hildebrandt}},
  \ and\ \bibinfo {author} {\bibnamefont {et~al.}},\ }\href {\doibase
  10.1093/mnras/stx1675} {\bibfield  {journal} {\bibinfo  {journal} {Monthly
  Notices of the Royal Astronomical Society}\ }\textbf {\bibinfo {volume}
  {471}},\ \bibinfo {pages} {1619–1633} (\bibinfo {year} {2017})}\BibitemShut
  {NoStop}%
\bibitem [{\citenamefont {Kirk}\ \emph {et~al.}(2016)\citenamefont {Kirk},
  \citenamefont {Omori}, \citenamefont {Benoit-Lévy}, \citenamefont {Cawthon},
  \citenamefont {Chang}, \citenamefont {Larsen}, \citenamefont {Amara},
  \citenamefont {Bacon}, \citenamefont {Crawford}, \citenamefont {Dodelson},\
  and\ \citenamefont {et~al.}}]{Kirk_2016}%
  \BibitemOpen
  \bibfield  {author} {\bibinfo {author} {\bibfnamefont {D.}~\bibnamefont
  {Kirk}}, \bibinfo {author} {\bibfnamefont {Y.}~\bibnamefont {Omori}},
  \bibinfo {author} {\bibfnamefont {A.}~\bibnamefont {Benoit-Lévy}}, \bibinfo
  {author} {\bibfnamefont {R.}~\bibnamefont {Cawthon}}, \bibinfo {author}
  {\bibfnamefont {C.}~\bibnamefont {Chang}}, \bibinfo {author} {\bibfnamefont
  {P.}~\bibnamefont {Larsen}}, \bibinfo {author} {\bibfnamefont
  {A.}~\bibnamefont {Amara}}, \bibinfo {author} {\bibfnamefont
  {D.}~\bibnamefont {Bacon}}, \bibinfo {author} {\bibfnamefont {T.~M.}\
  \bibnamefont {Crawford}}, \bibinfo {author} {\bibfnamefont {S.}~\bibnamefont
  {Dodelson}}, \ and\ \bibinfo {author} {\bibnamefont {et~al.}},\ }\href
  {\doibase 10.1093/mnras/stw570} {\bibfield  {journal} {\bibinfo  {journal}
  {Monthly Notices of the Royal Astronomical Society}\ }\textbf {\bibinfo
  {volume} {459}},\ \bibinfo {pages} {21–34} (\bibinfo {year}
  {2016})}\BibitemShut {NoStop}%
\bibitem [{\citenamefont {Liu}\ and\ \citenamefont {Hill}(2015)}]{Liu_2015}%
  \BibitemOpen
  \bibfield  {author} {\bibinfo {author} {\bibfnamefont {J.}~\bibnamefont
  {Liu}}\ and\ \bibinfo {author} {\bibfnamefont {J.~C.}\ \bibnamefont {Hill}},\
  }\href {\doibase 10.1103/physrevd.92.063517} {\bibfield  {journal} {\bibinfo
  {journal} {Physical Review D}\ }\textbf {\bibinfo {volume} {92}} (\bibinfo
  {year} {2015}),\ 10.1103/physrevd.92.063517}\BibitemShut {NoStop}%
\bibitem [{\citenamefont {Seljak}\ and\ \citenamefont
  {Zaldarriaga}(1999)}]{Seljak_1999}%
  \BibitemOpen
  \bibfield  {author} {\bibinfo {author} {\bibfnamefont {U.}~\bibnamefont
  {Seljak}}\ and\ \bibinfo {author} {\bibfnamefont {M.}~\bibnamefont
  {Zaldarriaga}},\ }\href {\doibase 10.1103/physrevd.60.043504} {\bibfield
  {journal} {\bibinfo  {journal} {Physical Review D}\ }\textbf {\bibinfo
  {volume} {60}} (\bibinfo {year} {1999}),\
  10.1103/physrevd.60.043504}\BibitemShut {NoStop}%
\bibitem [{\citenamefont {Goldberg}\ and\ \citenamefont
  {Spergel}(1999)}]{Goldberg_1999}%
  \BibitemOpen
  \bibfield  {author} {\bibinfo {author} {\bibfnamefont {D.~M.}\ \bibnamefont
  {Goldberg}}\ and\ \bibinfo {author} {\bibfnamefont {D.~N.}\ \bibnamefont
  {Spergel}},\ }\href {\doibase 10.1103/physrevd.59.103002} {\bibfield
  {journal} {\bibinfo  {journal} {Physical Review D}\ }\textbf {\bibinfo
  {volume} {59}} (\bibinfo {year} {1999}),\
  10.1103/physrevd.59.103002}\BibitemShut {NoStop}%
\bibitem [{\citenamefont {Namikawa}(2016)}]{Namikawa:2016jff}%
  \BibitemOpen
  \bibfield  {author} {\bibinfo {author} {\bibfnamefont {T.}~\bibnamefont
  {Namikawa}},\ }\href {\doibase 10.1103/PhysRevD.93.121301} {\bibfield
  {journal} {\bibinfo  {journal} {Phys. Rev. D}\ }\textbf {\bibinfo {volume}
  {93}},\ \bibinfo {pages} {121301} (\bibinfo {year} {2016})},\ \Eprint
  {http://arxiv.org/abs/1604.08578} {arXiv:1604.08578 [astro-ph.CO]}
  \BibitemShut {NoStop}%
\bibitem [{\citenamefont {Nguyen}\ \emph {et~al.}(2019)\citenamefont {Nguyen},
  \citenamefont {Sehgal},\ and\ \citenamefont
  {Madhavacheril}}]{Nguyen:2017zqu}%
  \BibitemOpen
  \bibfield  {author} {\bibinfo {author} {\bibfnamefont {H.~N.}\ \bibnamefont
  {Nguyen}}, \bibinfo {author} {\bibfnamefont {N.}~\bibnamefont {Sehgal}}, \
  and\ \bibinfo {author} {\bibfnamefont {M.}~\bibnamefont {Madhavacheril}},\
  }\href {\doibase 10.1103/PhysRevD.99.023502} {\bibfield  {journal} {\bibinfo
  {journal} {Phys. Rev. D}\ }\textbf {\bibinfo {volume} {99}},\ \bibinfo
  {pages} {023502} (\bibinfo {year} {2019})},\ \Eprint
  {http://arxiv.org/abs/1710.03747} {arXiv:1710.03747 [astro-ph.CO]}
  \BibitemShut {NoStop}%
\bibitem [{\citenamefont {Aguilar~Fa\'undez}\ \emph {et~al.}(2019)\citenamefont
  {Aguilar~Fa\'undez} \emph {et~al.}}]{Polarbear:2019zli}%
  \BibitemOpen
  \bibfield  {author} {\bibinfo {author} {\bibfnamefont {M.}~\bibnamefont
  {Aguilar~Fa\'undez}} \emph {et~al.} (\bibinfo {collaboration} {Polarbear}),\
  }\href {\doibase 10.3847/1538-4357/ab4a78} {\bibfield  {journal} {\bibinfo
  {journal} {Astrophys. J.}\ }\textbf {\bibinfo {volume} {886}},\ \bibinfo
  {pages} {38} (\bibinfo {year} {2019})},\ \Eprint
  {http://arxiv.org/abs/1903.07046} {arXiv:1903.07046 [astro-ph.CO]}
  \BibitemShut {NoStop}%
\bibitem [{\citenamefont {Das}\ \emph {et~al.}(2013)\citenamefont {Das},
  \citenamefont {Errard},\ and\ \citenamefont {Spergel}}]{Das:2013aia}%
  \BibitemOpen
  \bibfield  {author} {\bibinfo {author} {\bibfnamefont {S.}~\bibnamefont
  {Das}}, \bibinfo {author} {\bibfnamefont {J.}~\bibnamefont {Errard}}, \ and\
  \bibinfo {author} {\bibfnamefont {D.}~\bibnamefont {Spergel}},\ }\href@noop
  {} {\  (\bibinfo {year} {2013})},\ \Eprint {http://arxiv.org/abs/1311.2338}
  {arXiv:1311.2338 [astro-ph.CO]} \BibitemShut {NoStop}%
\bibitem [{\citenamefont {Larsen}\ and\ \citenamefont
  {Challinor}(2016)}]{Larsen:2015aoa}%
  \BibitemOpen
  \bibfield  {author} {\bibinfo {author} {\bibfnamefont {P.}~\bibnamefont
  {Larsen}}\ and\ \bibinfo {author} {\bibfnamefont {A.}~\bibnamefont
  {Challinor}},\ }\href {\doibase 10.1093/mnras/stw1645} {\bibfield  {journal}
  {\bibinfo  {journal} {Mon. Not. Roy. Astron. Soc.}\ }\textbf {\bibinfo
  {volume} {461}},\ \bibinfo {pages} {4343} (\bibinfo {year} {2016})},\ \Eprint
  {http://arxiv.org/abs/1510.02617} {arXiv:1510.02617 [astro-ph.CO]}
  \BibitemShut {NoStop}%
\bibitem [{\citenamefont {{Schaan}}\ \emph {et~al.}(2017)\citenamefont
  {{Schaan}}, \citenamefont {{Krause}}, \citenamefont {{Eifler}}, \citenamefont
  {{Dor{\'e}}}, \citenamefont {{Miyatake}}, \citenamefont {{Rhodes}},\ and\
  \citenamefont {{Spergel}}}]{2017PhRvD..95l3512S}%
  \BibitemOpen
  \bibfield  {author} {\bibinfo {author} {\bibfnamefont {E.}~\bibnamefont
  {{Schaan}}}, \bibinfo {author} {\bibfnamefont {E.}~\bibnamefont {{Krause}}},
  \bibinfo {author} {\bibfnamefont {T.}~\bibnamefont {{Eifler}}}, \bibinfo
  {author} {\bibfnamefont {O.}~\bibnamefont {{Dor{\'e}}}}, \bibinfo {author}
  {\bibfnamefont {H.}~\bibnamefont {{Miyatake}}}, \bibinfo {author}
  {\bibfnamefont {J.}~\bibnamefont {{Rhodes}}}, \ and\ \bibinfo {author}
  {\bibfnamefont {D.~N.}\ \bibnamefont {{Spergel}}},\ }\href {\doibase
  10.1103/PhysRevD.95.123512} {\bibfield  {journal} {\bibinfo  {journal}
  {\prd}\ }\textbf {\bibinfo {volume} {95}},\ \bibinfo {eid} {123512} (\bibinfo
  {year} {2017})},\ \Eprint {http://arxiv.org/abs/1607.01761} {arXiv:1607.01761
  [astro-ph.CO]} \BibitemShut {NoStop}%
\bibitem [{\citenamefont {{Vallinotto}}(2012)}]{2012ApJ...759...32V}%
  \BibitemOpen
  \bibfield  {author} {\bibinfo {author} {\bibfnamefont {A.}~\bibnamefont
  {{Vallinotto}}},\ }\href {\doibase 10.1088/0004-637X/759/1/32} {\bibfield
  {journal} {\bibinfo  {journal} {\apj}\ }\textbf {\bibinfo {volume} {759}},\
  \bibinfo {eid} {32} (\bibinfo {year} {2012})},\ \Eprint
  {http://arxiv.org/abs/1110.5339} {arXiv:1110.5339 [astro-ph.CO]} \BibitemShut
  {NoStop}%
\bibitem [{\citenamefont {Okamoto}\ and\ \citenamefont
  {Hu}(2003)}]{Okamoto:2003zw}%
  \BibitemOpen
  \bibfield  {author} {\bibinfo {author} {\bibfnamefont {T.}~\bibnamefont
  {Okamoto}}\ and\ \bibinfo {author} {\bibfnamefont {W.}~\bibnamefont {Hu}},\
  }\href {\doibase 10.1103/PhysRevD.67.083002} {\bibfield  {journal} {\bibinfo
  {journal} {Phys. Rev. D}\ }\textbf {\bibinfo {volume} {67}},\ \bibinfo
  {pages} {083002} (\bibinfo {year} {2003})},\ \Eprint
  {http://arxiv.org/abs/astro-ph/0301031} {arXiv:astro-ph/0301031} \BibitemShut
  {NoStop}%
\bibitem [{\citenamefont {Krause}\ and\ \citenamefont
  {Hirata}(2010)}]{Krause:2009yr}%
  \BibitemOpen
  \bibfield  {author} {\bibinfo {author} {\bibfnamefont {E.}~\bibnamefont
  {Krause}}\ and\ \bibinfo {author} {\bibfnamefont {C.~M.}\ \bibnamefont
  {Hirata}},\ }\href {\doibase 10.1051/0004-6361/200913524} {\bibfield
  {journal} {\bibinfo  {journal} {Astron. Astrophys.}\ }\textbf {\bibinfo
  {volume} {523}},\ \bibinfo {pages} {A28} (\bibinfo {year} {2010})},\ \Eprint
  {http://arxiv.org/abs/0910.3786} {arXiv:0910.3786 [astro-ph.CO]} \BibitemShut
  {NoStop}%
\bibitem [{\citenamefont {Pratten}\ and\ \citenamefont
  {Lewis}(2016)}]{Pratten:2016dsm}%
  \BibitemOpen
  \bibfield  {author} {\bibinfo {author} {\bibfnamefont {G.}~\bibnamefont
  {Pratten}}\ and\ \bibinfo {author} {\bibfnamefont {A.}~\bibnamefont
  {Lewis}},\ }\href {\doibase 10.1088/1475-7516/2016/08/047} {\bibfield
  {journal} {\bibinfo  {journal} {JCAP}\ }\textbf {\bibinfo {volume} {1608}},\
  \bibinfo {pages} {047} (\bibinfo {year} {2016})},\ \Eprint
  {http://arxiv.org/abs/1605.05662} {arXiv:1605.05662 [astro-ph.CO]}
  \BibitemShut {NoStop}%
\bibitem [{\citenamefont {Fanizza}\ \emph {et~al.}(2015)\citenamefont
  {Fanizza}, \citenamefont {Gasperini}, \citenamefont {Marozzi},\ and\
  \citenamefont {Veneziano}}]{Fanizza:2015swa}%
  \BibitemOpen
  \bibfield  {author} {\bibinfo {author} {\bibfnamefont {G.}~\bibnamefont
  {Fanizza}}, \bibinfo {author} {\bibfnamefont {M.}~\bibnamefont {Gasperini}},
  \bibinfo {author} {\bibfnamefont {G.}~\bibnamefont {Marozzi}}, \ and\
  \bibinfo {author} {\bibfnamefont {G.}~\bibnamefont {Veneziano}},\ }\href
  {\doibase 10.1088/1475-7516/2015/08/020} {\bibfield  {journal} {\bibinfo
  {journal} {JCAP}\ }\textbf {\bibinfo {volume} {08}},\ \bibinfo {pages} {020}
  (\bibinfo {year} {2015})},\ \Eprint {http://arxiv.org/abs/1506.02003}
  {arXiv:1506.02003 [astro-ph.CO]} \BibitemShut {NoStop}%
\bibitem [{\citenamefont {Marozzi}\ \emph {et~al.}(2016)\citenamefont
  {Marozzi}, \citenamefont {Fanizza}, \citenamefont {Di~Dio},\ and\
  \citenamefont {Durrer}}]{Marozzi:2016uob}%
  \BibitemOpen
  \bibfield  {author} {\bibinfo {author} {\bibfnamefont {G.}~\bibnamefont
  {Marozzi}}, \bibinfo {author} {\bibfnamefont {G.}~\bibnamefont {Fanizza}},
  \bibinfo {author} {\bibfnamefont {E.}~\bibnamefont {Di~Dio}}, \ and\ \bibinfo
  {author} {\bibfnamefont {R.}~\bibnamefont {Durrer}},\ }\href {\doibase
  10.1088/1475-7516/2016/09/028} {\bibfield  {journal} {\bibinfo  {journal}
  {JCAP}\ }\textbf {\bibinfo {volume} {09}},\ \bibinfo {pages} {028} (\bibinfo
  {year} {2016})},\ \Eprint {http://arxiv.org/abs/1605.08761} {arXiv:1605.08761
  [astro-ph.CO]} \BibitemShut {NoStop}%
\bibitem [{\citenamefont {Fabbian}\ \emph {et~al.}(2019)\citenamefont
  {Fabbian}, \citenamefont {Lewis},\ and\ \citenamefont
  {Beck}}]{Fabbian:2019tik}%
  \BibitemOpen
  \bibfield  {author} {\bibinfo {author} {\bibfnamefont {G.}~\bibnamefont
  {Fabbian}}, \bibinfo {author} {\bibfnamefont {A.}~\bibnamefont {Lewis}}, \
  and\ \bibinfo {author} {\bibfnamefont {D.}~\bibnamefont {Beck}},\ }\href
  {\doibase 10.1088/1475-7516/2019/10/057} {\bibfield  {journal} {\bibinfo
  {journal} {JCAP}\ }\textbf {\bibinfo {volume} {10}},\ \bibinfo {pages} {057}
  (\bibinfo {year} {2019})},\ \Eprint {http://arxiv.org/abs/1906.08760}
  {arXiv:1906.08760 [astro-ph.CO]} \BibitemShut {NoStop}%
\bibitem [{\citenamefont {Lewis}\ and\ \citenamefont
  {Challinor}(2006)}]{Lewis:2006fu}%
  \BibitemOpen
  \bibfield  {author} {\bibinfo {author} {\bibfnamefont {A.}~\bibnamefont
  {Lewis}}\ and\ \bibinfo {author} {\bibfnamefont {A.}~\bibnamefont
  {Challinor}},\ }\href {\doibase 10.1016/j.physrep.2006.03.002} {\bibfield
  {journal} {\bibinfo  {journal} {Phys. Rept.}\ }\textbf {\bibinfo {volume}
  {429}},\ \bibinfo {pages} {1} (\bibinfo {year} {2006})},\ \Eprint
  {http://arxiv.org/abs/astro-ph/0601594} {arXiv:astro-ph/0601594 [astro-ph]}
  \BibitemShut {NoStop}%
\bibitem [{\citenamefont {Marozzi}\ \emph {et~al.}(2017)\citenamefont
  {Marozzi}, \citenamefont {Fanizza}, \citenamefont {Di~Dio},\ and\
  \citenamefont {Durrer}}]{Marozzi:2016und}%
  \BibitemOpen
  \bibfield  {author} {\bibinfo {author} {\bibfnamefont {G.}~\bibnamefont
  {Marozzi}}, \bibinfo {author} {\bibfnamefont {G.}~\bibnamefont {Fanizza}},
  \bibinfo {author} {\bibfnamefont {E.}~\bibnamefont {Di~Dio}}, \ and\ \bibinfo
  {author} {\bibfnamefont {R.}~\bibnamefont {Durrer}},\ }\href {\doibase
  10.1103/PhysRevLett.118.211301} {\bibfield  {journal} {\bibinfo  {journal}
  {Phys. Rev. Lett.}\ }\textbf {\bibinfo {volume} {118}},\ \bibinfo {pages}
  {211301} (\bibinfo {year} {2017})},\ \Eprint
  {http://arxiv.org/abs/1612.07650} {arXiv:1612.07650 [astro-ph.CO]}
  \BibitemShut {NoStop}%
\bibitem [{\citenamefont {Fabbian}\ \emph {et~al.}(2018)\citenamefont
  {Fabbian}, \citenamefont {Calabrese},\ and\ \citenamefont
  {Carbone}}]{Fabbian:2017wfp}%
  \BibitemOpen
  \bibfield  {author} {\bibinfo {author} {\bibfnamefont {G.}~\bibnamefont
  {Fabbian}}, \bibinfo {author} {\bibfnamefont {M.}~\bibnamefont {Calabrese}},
  \ and\ \bibinfo {author} {\bibfnamefont {C.}~\bibnamefont {Carbone}},\ }\href
  {\doibase 10.1088/1475-7516/2018/02/050} {\bibfield  {journal} {\bibinfo
  {journal} {JCAP}\ }\textbf {\bibinfo {volume} {02}},\ \bibinfo {pages} {050}
  (\bibinfo {year} {2018})},\ \Eprint {http://arxiv.org/abs/1702.03317}
  {arXiv:1702.03317 [astro-ph.CO]} \BibitemShut {NoStop}%
\bibitem [{\citenamefont {Pitrou}(2009)}]{Pitrou:2008hy}%
  \BibitemOpen
  \bibfield  {author} {\bibinfo {author} {\bibfnamefont {C.}~\bibnamefont
  {Pitrou}},\ }\href {\doibase 10.1088/0264-9381/26/6/065006} {\bibfield
  {journal} {\bibinfo  {journal} {Class. Quant. Grav.}\ }\textbf {\bibinfo
  {volume} {26}},\ \bibinfo {pages} {065006} (\bibinfo {year} {2009})},\
  \Eprint {http://arxiv.org/abs/0809.3036} {arXiv:0809.3036 [gr-qc]}
  \BibitemShut {NoStop}%
\bibitem [{\citenamefont {Fanizza}\ \emph {et~al.}(2018)\citenamefont
  {Fanizza}, \citenamefont {Yoo},\ and\ \citenamefont
  {Biern}}]{Fanizza:2018qux}%
  \BibitemOpen
  \bibfield  {author} {\bibinfo {author} {\bibfnamefont {G.}~\bibnamefont
  {Fanizza}}, \bibinfo {author} {\bibfnamefont {J.}~\bibnamefont {Yoo}}, \ and\
  \bibinfo {author} {\bibfnamefont {S.~G.}\ \bibnamefont {Biern}},\ }\href
  {\doibase 10.1088/1475-7516/2018/09/037} {\bibfield  {journal} {\bibinfo
  {journal} {JCAP}\ }\textbf {\bibinfo {volume} {09}},\ \bibinfo {pages} {037}
  (\bibinfo {year} {2018})},\ \Eprint {http://arxiv.org/abs/1805.05959}
  {arXiv:1805.05959 [gr-qc]} \BibitemShut {NoStop}%
\bibitem [{\citenamefont {Seljak}\ and\ \citenamefont
  {Zaldarriaga}(1996)}]{Seljak:1996is}%
  \BibitemOpen
  \bibfield  {author} {\bibinfo {author} {\bibfnamefont {U.}~\bibnamefont
  {Seljak}}\ and\ \bibinfo {author} {\bibfnamefont {M.}~\bibnamefont
  {Zaldarriaga}},\ }\href {\doibase 10.1086/177793} {\bibfield  {journal}
  {\bibinfo  {journal} {Astrophys. J.}\ }\textbf {\bibinfo {volume} {469}},\
  \bibinfo {pages} {437} (\bibinfo {year} {1996})},\ \Eprint
  {http://arxiv.org/abs/astro-ph/9603033} {arXiv:astro-ph/9603033 [astro-ph]}
  \BibitemShut {NoStop}%
\bibitem [{\citenamefont {Dodelson}(2003)}]{Dodelson:2003ft}%
  \BibitemOpen
  \bibfield  {author} {\bibinfo {author} {\bibfnamefont {S.}~\bibnamefont
  {Dodelson}},\ }\href
  {http://www.slac.stanford.edu/spires/find/books/www?cl=QB981:D62:2003} {\emph
  {\bibinfo {title} {{Modern Cosmology}}}}\ (\bibinfo  {publisher} {Academic
  Press},\ \bibinfo {address} {Amsterdam},\ \bibinfo {year} {2003})\BibitemShut
  {NoStop}%
\bibitem [{\citenamefont {Calabrese}\ \emph {et~al.}(2015)\citenamefont
  {Calabrese}, \citenamefont {Carbone}, \citenamefont {Fabbian}, \citenamefont
  {Baldi},\ and\ \citenamefont {Baccigalupi}}]{Calabrese:2014gla}%
  \BibitemOpen
  \bibfield  {author} {\bibinfo {author} {\bibfnamefont {M.}~\bibnamefont
  {Calabrese}}, \bibinfo {author} {\bibfnamefont {C.}~\bibnamefont {Carbone}},
  \bibinfo {author} {\bibfnamefont {G.}~\bibnamefont {Fabbian}}, \bibinfo
  {author} {\bibfnamefont {M.}~\bibnamefont {Baldi}}, \ and\ \bibinfo {author}
  {\bibfnamefont {C.}~\bibnamefont {Baccigalupi}},\ }\href {\doibase
  10.1088/1475-7516/2015/03/049} {\bibfield  {journal} {\bibinfo  {journal}
  {JCAP}\ }\textbf {\bibinfo {volume} {03}},\ \bibinfo {pages} {049} (\bibinfo
  {year} {2015})},\ \Eprint {http://arxiv.org/abs/1409.7680} {arXiv:1409.7680
  [astro-ph.CO]} \BibitemShut {NoStop}%
\bibitem [{\citenamefont {Petri}\ \emph {et~al.}(2017)\citenamefont {Petri},
  \citenamefont {Haiman},\ and\ \citenamefont {May}}]{Petri:2016qya}%
  \BibitemOpen
  \bibfield  {author} {\bibinfo {author} {\bibfnamefont {A.}~\bibnamefont
  {Petri}}, \bibinfo {author} {\bibfnamefont {Z.}~\bibnamefont {Haiman}}, \
  and\ \bibinfo {author} {\bibfnamefont {M.}~\bibnamefont {May}},\ }\href
  {\doibase 10.1103/PhysRevD.95.123503} {\bibfield  {journal} {\bibinfo
  {journal} {Phys. Rev. D}\ }\textbf {\bibinfo {volume} {95}},\ \bibinfo
  {pages} {123503} (\bibinfo {year} {2017})},\ \Eprint
  {http://arxiv.org/abs/1612.00852} {arXiv:1612.00852 [astro-ph.CO]}
  \BibitemShut {NoStop}%
\bibitem [{\citenamefont {Barthelemy}\ \emph {et~al.}(2020)\citenamefont
  {Barthelemy}, \citenamefont {Codis},\ and\ \citenamefont
  {Bernardeau}}]{Barthelemy:2020igw}%
  \BibitemOpen
  \bibfield  {author} {\bibinfo {author} {\bibfnamefont {A.}~\bibnamefont
  {Barthelemy}}, \bibinfo {author} {\bibfnamefont {S.}~\bibnamefont {Codis}}, \
  and\ \bibinfo {author} {\bibfnamefont {F.}~\bibnamefont {Bernardeau}},\
  }\href {\doibase 10.1093/mnras/staa931} {\bibfield  {journal} {\bibinfo
  {journal} {Mon. Not. Roy. Astron. Soc.}\ }\textbf {\bibinfo {volume} {494}},\
  \bibinfo {pages} {3368} (\bibinfo {year} {2020})},\ \Eprint
  {http://arxiv.org/abs/2002.03625} {arXiv:2002.03625 [astro-ph.CO]}
  \BibitemShut {NoStop}%
\bibitem [{\citenamefont {Shapiro}\ and\ \citenamefont
  {Cooray}(2006)}]{Shapiro:2006em}%
  \BibitemOpen
  \bibfield  {author} {\bibinfo {author} {\bibfnamefont {C.}~\bibnamefont
  {Shapiro}}\ and\ \bibinfo {author} {\bibfnamefont {A.}~\bibnamefont
  {Cooray}},\ }\href {\doibase 10.1088/1475-7516/2006/03/007} {\bibfield
  {journal} {\bibinfo  {journal} {JCAP}\ }\textbf {\bibinfo {volume} {0603}},\
  \bibinfo {pages} {007} (\bibinfo {year} {2006})},\ \Eprint
  {http://arxiv.org/abs/astro-ph/0601226} {arXiv:astro-ph/0601226 [astro-ph]}
  \BibitemShut {NoStop}%
\bibitem [{\citenamefont {Beck}\ \emph {et~al.}(2018)\citenamefont {Beck},
  \citenamefont {Fabbian},\ and\ \citenamefont {Errard}}]{Beck:2018wud}%
  \BibitemOpen
  \bibfield  {author} {\bibinfo {author} {\bibfnamefont {D.}~\bibnamefont
  {Beck}}, \bibinfo {author} {\bibfnamefont {G.}~\bibnamefont {Fabbian}}, \
  and\ \bibinfo {author} {\bibfnamefont {J.}~\bibnamefont {Errard}},\ }\href
  {\doibase 10.1103/PhysRevD.98.043512} {\bibfield  {journal} {\bibinfo
  {journal} {Phys. Rev. D}\ }\textbf {\bibinfo {volume} {98}},\ \bibinfo
  {pages} {043512} (\bibinfo {year} {2018})},\ \Eprint
  {http://arxiv.org/abs/1806.01216} {arXiv:1806.01216 [astro-ph.CO]}
  \BibitemShut {NoStop}%
\bibitem [{\citenamefont {Takahashi}\ \emph {et~al.}(2017)\citenamefont
  {Takahashi}, \citenamefont {Hamana}, \citenamefont {Shirasaki}, \citenamefont
  {Namikawa}, \citenamefont {Nishimichi}, \citenamefont {Osato},\ and\
  \citenamefont {Shiroyama}}]{Takahashi:2017hjr}%
  \BibitemOpen
  \bibfield  {author} {\bibinfo {author} {\bibfnamefont {R.}~\bibnamefont
  {Takahashi}}, \bibinfo {author} {\bibfnamefont {T.}~\bibnamefont {Hamana}},
  \bibinfo {author} {\bibfnamefont {M.}~\bibnamefont {Shirasaki}}, \bibinfo
  {author} {\bibfnamefont {T.}~\bibnamefont {Namikawa}}, \bibinfo {author}
  {\bibfnamefont {T.}~\bibnamefont {Nishimichi}}, \bibinfo {author}
  {\bibfnamefont {K.}~\bibnamefont {Osato}}, \ and\ \bibinfo {author}
  {\bibfnamefont {K.}~\bibnamefont {Shiroyama}},\ }\href {\doibase
  10.3847/1538-4357/aa943d} {\bibfield  {journal} {\bibinfo  {journal}
  {Astrophys. J.}\ }\textbf {\bibinfo {volume} {850}},\ \bibinfo {pages} {24}
  (\bibinfo {year} {2017})},\ \Eprint {http://arxiv.org/abs/1706.01472}
  {arXiv:1706.01472 [astro-ph.CO]} \BibitemShut {NoStop}%
\bibitem [{\citenamefont {B\"ohm}\ \emph {et~al.}(2020)\citenamefont {B\"ohm},
  \citenamefont {Modi},\ and\ \citenamefont {Castorina}}]{Bohm:2019bek}%
  \BibitemOpen
  \bibfield  {author} {\bibinfo {author} {\bibfnamefont {V.}~\bibnamefont
  {B\"ohm}}, \bibinfo {author} {\bibfnamefont {C.}~\bibnamefont {Modi}}, \ and\
  \bibinfo {author} {\bibfnamefont {E.}~\bibnamefont {Castorina}},\ }\href
  {\doibase 10.1088/1475-7516/2020/03/045} {\bibfield  {journal} {\bibinfo
  {journal} {JCAP}\ }\textbf {\bibinfo {volume} {03}},\ \bibinfo {pages} {045}
  (\bibinfo {year} {2020})},\ \Eprint {http://arxiv.org/abs/1910.06722}
  {arXiv:1910.06722 [astro-ph.CO]} \BibitemShut {NoStop}%
\end{thebibliography}%
\appendix
\section{Diagrammatic Representation for Remapping Ansatz} \label{sect:diagramrules}
In this appendix, we describe the diagrammatic rules for the construction of deflection integrals to all orders following the remapping \emph{ansatz}, and use this formalism to prove that the number of diagrams at $N$-th order is given by the Catalan number. We will drop $\nh$ from the following for notational simplicity.  In this section, ``order $N$'' refers to the order of the deflection $\delta x^a{}^{[N]}$. We emphasise that this representation is different to the diagrammatic representation for the Boltzmann formalism we introduced in \cite{Su:2014mga}.

The basic archetype is the dashed line connecting a left node associated with position $r$ to a right node with position $r_1$ which corresponds to an $\int dr_1$ integral over the kernel $W(r,r_1)$
\begin{equation}
\begin{tikzpicture}[baseline=(current bounding box.west)]
\fill (0,0) circle[radius=\dotsize]; \draw (0,0) node[below]{$r$};
\fill (1.2,0) \Square{\dotsize}; \draw (1.2,0) node[below]{$M^{a_1a_2\dots a_i}(r_1)$};
\draw[dashed] (0,0) to [out=60, in=120] (1.2,0);
\end{tikzpicture}
\equiv -2r\int_0^rdr_1 W(r_1,r)r_1\partial^a\chi(r_1)~,
\label{eqn:basic_1}
\end{equation}
where $M^{a_1a_2\dots a_i}$ is an object with $i$ indices constructed out of products  of $\delta x^a$. $\chi(r_1)$ is an unindexed object which is a full contraction of $M^{a_1a_2\dots a_i}$ with the $i$-th order term of the Taylor expansion of $\Psi_W(\bar{\bx} + \delta \bx)$ which are products of partial derivatives on $\Psi_W$, e.g. 
\begin{equation*}
\chi \in \frac{1}{i!}\partial_{a_1}\partial_{a_2}\dots\partial_{a_i}\Psi_W\delta x^{a_1}\delta x^{a_2}\dots\delta x^{a_i}~,~ \forall i\geq 0~.
\end{equation*}
At $i=0$, $\chi = \Psi_W$. $\delta x^a$ itself can be further expanded as $\delta x^a = \delta x^a{}^{[I]} + \delta x^a{}^{[II]} + \dots$, though the structure remains the same. In words, we say that ``the dashed line is acting on object $M^{a_1a_2\dots a_i}$''.  Thus the dashed line represents the Green's function solution of a source $\partial^a \chi$.

Using this archetype, and starting from the lowest order term, we introduce two further rules that allow for the consistent construction of all higher order terms.  The lowest order term is the first order deflection when the dashed line acts on the Weyl potential $\chi=\Psi_W$,
\begin{equation} 
\delta x^a{}^{[I]}=
\begin{tikzpicture}[baseline=(current bounding box.west)]
\fill (0,0) circle[radius=\dotsize]; \draw (0,0) node[below]{$r$};
\fill (0.6,0) \Square{\dotsize}; \draw (0.6,0) node[below]{$r_1$};
\draw[dashed] (0,0) to [out=60, in=120] (0.6,0);
\end{tikzpicture}
=-2r\int_0^rdr_1 W(r_1,r)r_1\partial^a \Psi_W(r_1)~,
\label{eqn:basic_2}
\end{equation}
thus $i=0$ since $\Psi_W$ has no index.

\emph{Rule 1: Linking rule.} The action of a dashed line on $M$ with $i>0$ is to \emph{link} the diagram of $M$ at the node associated with its $r$ argument (the left-most node).  Consider the action on a single indexed object $i=1$. At lowest order, this is $\delta x^b{}^{[I]}$, then $\chi = \partial_b \Psi_W \delta x^b{}^{[I]}$, and thus diagrammatically
\begin{eqnarray}
\begin{tikzpicture}[baseline=(current bounding box.west)]
\fill (0,0) circle[radius=\dotsize]; \draw (0,0) node[below]{$r$};
\fill (1,0) \Square{\dotsize}; \draw (1,0) node[below]{$\delta x^b{}^{[I]}(r_1)$};
\draw[dashed] (0,0) to [out=60, in=120] (1.0,0);
\end{tikzpicture} 
&=& -2r\int^r_0 dr_1 W(r_1,r)r_1\partial^a\partial_b\Psi_W(r_1)\underbrace{\delta x^b{}^{[I]}(r_1)}_{
\begin{tikzpicture}[baseline=(current bounding box.west)]
\fill (0,0) circle[radius=\dotsize]; \draw (0,0) node[below]{$r_1$};
\fill (0.6,0) \Square{\dotsize}; \draw (0.6,0) node[below]{$r_2$};
\draw[dashed] (0,0) to [out=60, in=120] (0.6,0);
\end{tikzpicture}
} \nn
&= &
\begin{tikzpicture}[baseline=(current bounding box.west)]
\fill (0,0) circle[radius=\dotsize]; \draw (0,0) node[below]{$r$};
\fill (0.6,0) \Square{\dotsize}; \draw (0.6,0) node[below]{$r_1$};
\fill (1.2,0) \Square{\dotsize}; \draw (1.2,0) node[below]{$r_2$};
\draw[dashed] (0,0) to [out=60, in=120] (0.6,0);
\draw[dashed] (0.6,0) to [out=60, in=120] (1.2,0);
\end{tikzpicture} \nn
&=&\delta x^a{}^{[II]}(r)~,  
\label{eqn:basic_3}
\end{eqnarray}
where in the 2nd line, our formalism has allow us to replace
\begin{tikzpicture}[baseline=(current bounding box.west)]
\fill (0,0) \Square{\dotsize}; \draw (0,0) node[below]{$\delta x^a{}^{[I]}(r_1)$};
\end{tikzpicture} 
with its diagrammatic representation \begin{tikzpicture}[baseline=(current bounding box.west)]
\fill (0,0) circle[radius=\dotsize]; \draw (0,0) node[below]{$r_1$};
\fill (0.6,0) \Square{\dotsize}; \draw (0.6,0) node[below]{$r_2$};
\draw[dashed] (0,0) to [out=60, in=120] (0.6,0);
\end{tikzpicture}
 where we have chosen $r_2$ to be its integration variable\footnote{It is convenient to choose integration variables such that the time ordering $r_i > r_{i+1}$ is obeyed.}. 

For an object with more than one  linked node, the linking rule will link to the node associated with its $r$ argument, for example, 
\begin{equation}
\begin{tikzpicture}[baseline=(current bounding box.west)]
\fill (0,0) circle[radius=\dotsize]; \draw (0,0) node[below]{$r$};
\fill (1,0) \Square{\dotsize}; \draw (1,0) node[below]{$\delta x^b{}^{[II]}(r_1)$};
\draw[dashed] (0,0) to [out=60, in=120] (1.0,0);
\end{tikzpicture} 
=
\begin{tikzpicture}[baseline=(current bounding box.west)]
\fill (0,0) circle[radius=\dotsize]; \draw (0,0) node[below]{$r$};
\fill (0.6,0) \Square{\dotsize}; \draw (0.6,0) node[below]{$r_1$};
\fill (1.2,0) \Square{\dotsize}; \draw (1.2,0) node[below]{$r_2$};
\fill (1.8,0) \Square{\dotsize}; \draw (1.8,0) node[below]{$r_3$};
\draw[dashed] (0,0) to [out=60, in=120] (0.6,0);
\draw[dashed] (0.6,0) to [out=60, in=120] (1.2,0);
\draw[dashed] (1.2,0) to [out=60, in=120] (1.8,0);
\end{tikzpicture}
\label{eqn:not_deltaIII}
\end{equation}
and so on (although note that \eqn{eqn:not_deltaIII} is only one of the two terms of $\delta x^a{}^{[III]}$-- see \eqn{eqn:deltaIII}).

\emph{Rule 2: Product rule.} Objects with index $i>1$ are products of $\delta x^{a_1}{}^{[N_1]}$, $\delta x^{a_2}{}^{[N_2]}$, $\dots$, etc. The order of this object is then the sum $N= \sum_i N_i$.  Since they all have the same argument, we link to \emph{all} of them. Diagrammatically, we represent them with a product symbol ``$\times$''
\begin{equation}
\begin{tikzpicture}[baseline=(current bounding box.west)]
\fill (0,0) \Square{\dotsize}; \draw (0,0) node[below]{$M^{a_1a_2\dots }(r)$};
\end{tikzpicture}
= 
\begin{tikzpicture}[baseline=(current bounding box.west)]
\fill (0,0) \Square{\dotsize}; \draw (0,0) node[below]{$\delta x^{a_1}{}^{[N_1]}(r_1)$};
\end{tikzpicture}
\times
\begin{tikzpicture}[baseline=(current bounding box.west)]
\fill (0,0) \Square{\dotsize}; \draw (0,0) node[below]{$\delta x^{a_2}{}^{[N_2]}(r_1)$};
\end{tikzpicture}
\times \dots 
\end{equation}
Since $M$, and hence all the $\delta x^a$'s are evaluated at the same point $r$, the action of the dashed line on $M$ must integrate over all of them. Thus we represent it as multiple dashed lines terminating at the same node associated with $r$,
\begin{equation}
\begin{tikzpicture}[baseline=(current bounding box.west)]
\fill (0,0) circle[radius=\dotsize]; \draw (0,0) node[below]{$r$};
\fill (1.4,0) \Square{\dotsize}; \draw (1.4,0) node[below]{$M^{a_1a_2\dots}(r_1)$};
\draw[dashed] (0,0) to [out=60, in=120] (1.4,0);
\end{tikzpicture}
= 
\begin{tikzpicture}[baseline=(current bounding box.west)]
\fill (0,0) circle[radius=\dotsize]; \draw (0,0) node[below]{$r$};
\fill (1,0) \Square{\dotsize}; \draw (1,0) node[below]{$\delta x^{a_1}{}^{[N_1]}(r_1)$};
\draw (2,0)  node{$\times$};
\fill (3,0) \Square{\dotsize}; \draw (3,0) node[below]{$\delta x^{a_2}{}^{[N_2]}(r_1)$};
\draw (4,0)  node{$\times$};
\draw (5,0)  node{$\dots$};
\draw[dashed] (0,0) to [out=60, in=120] (1,0);
\draw[dashed] (0,0) to [out=60, in=120] (3,0);
\end{tikzpicture}
\label{eqn:basic_4}
\end{equation}
We can then reconstruct the full diagram by replacing the $\delta x^a{}^{[N_i]}$ nodes with their corresponding diagrams, using the linking rule, for example for $M^{ab} =\delta x^{a}{}^{[I]}\delta x^{a}{}^{[II]}$, we get
\begin{eqnarray}
\begin{tikzpicture}[baseline=(current bounding box.west)]
\fill (0,0) circle[radius=\dotsize]; \draw (0,0) node[below]{$r$};
\fill (1.4,0) \Square{\dotsize}; \draw (1.4,0) node[below]{$\delta x^{a}{}^{[I]}\delta x^{a}{}^{[II]}$};
\draw[dashed] (0,0) to [out=60, in=120] (1.4,0);
\end{tikzpicture}
&=
\begin{tikzpicture}[baseline=(current bounding box.west)]
\fill (0,0) circle[radius=\dotsize]; \draw (0,0) node[below]{$r$};
\fill (1,0) \Square{\dotsize}; \draw (1,0) node[below]{$\delta x^{a}{}^{[I]}(r_1)$};
\draw (2,0)  node{$\times$};
\fill (3,0) \Square{\dotsize}; \draw (3,0) node[below]{$\delta x^{b}{}^{[II]}(r_1)$};
\draw[dashed] (0,0) to [out=60, in=120] (1,0);
\draw[dashed] (0,0) to [out=60, in=120] (3,0);
\end{tikzpicture}
\nn
&=
\begin{tikzpicture}[baseline=(current bounding box.west)]
\fill (0,0) circle[radius=\dotsize]; \draw (0,0) node[below]{$r$};
\fill (1,0) \Square{\dotsize}; \draw (1,0) node[below]{$r_1$};
\fill (1.6,0) \Square{\dotsize}; \draw (1.6,0) node[below]{$r_2$};
\draw[dashed] (1,0) to [out=60, in=120] (1.6,0);
\draw (2.3,0)  node{$\times$};
\fill (3,0) \Square{\dotsize}; \draw (3,0) node[below]{$r_1$};
\fill (3.6,0) \Square{\dotsize}; \draw (3.6,0) node[below]{$r_3$};
\fill (4.2,0) \Square{\dotsize}; \draw (4.2,0) node[below]{$r_4$};
\draw[dashed] (3,0) to [out=60, in=120] (3.6,0);
\draw[dashed] (3.6,0) to [out=60, in=120] (4.2,0);
\draw[dashed] (0,0) to [out=60, in=120] (1,0);
\draw[dashed] (0,0) to [out=60, in=120] (3,0);
\end{tikzpicture}
\label{eqn:basic_5}
\end{eqnarray}
where we have used the archetypes \eqn{eqn:basic_2} and \eqn{eqn:basic_3} in the 2nd line. Finally, without any ambiguity, we can combine the $r_1$ nodes into a single node as follows
\begin{equation}
\begin{tikzpicture}[baseline=(current bounding box.west)]
\fill (0,0) \Square{\dotsize}; \draw (0,0) node[below]{$r_1$};
\fill (0.6,0) \Square{\dotsize}; \draw (0.6,0) node[below]{$r_2$};
\draw[dashed] (0,0) to [out=60, in=120] (0.6,0);
\draw (1.2,0)  node{$\times$};
\fill (1.8,0) \Square{\dotsize}; \draw (1.8,0) node[below]{$r_1$};
\fill (2.4,0) \Square{\dotsize}; \draw (2.4,0) node[below]{$r_3$};
\fill (3.0,0) \Square{\dotsize}; \draw (3.0,0) node[below]{$r_4$};
\draw[dashed] (1.8,0) to [out=60, in=120] (2.4,0);
\draw[dashed] (2.4,0) to [out=60, in=120] (3.0,0);
\end{tikzpicture}
=
\begin{tikzpicture}[baseline=(current bounding box.west)]
\fill (0,0) \Square{\dotsize}; \draw (0,0) node[below]{$r_1$};
\fill (0.6,0) \Square{\dotsize}; \draw (0.6,0) node[below]{$r_2$};
\fill (1.2,0) \Square{\dotsize}; \draw (1.2,0) node[below]{$r_3$};
\fill (1.8,0) \Square{\dotsize}; \draw (1.8,0) node[below]{$r_4$};
\draw[dashed] (0,0) to [out=60, in=120] (0.6,0);
\draw[dashed] (0,0) to [out=60, in=120] (1.2,0);
\draw[dashed] (1.2,0) to [out=60, in=120] (1.8,0);
\end{tikzpicture}
\end{equation}
and thus \eqn{eqn:basic_5} becomes
\begin{equation}
\begin{tikzpicture}[baseline=(current bounding box.west)]
\fill (0,0) circle[radius=\dotsize]; \draw (0,0) node[below]{$r$};
\fill (1.4,0) \Square{\dotsize}; \draw (1.4,0) node[below]{$\delta x^{a}{}^{[I]}\delta x^{a}{}^{[II]}$};
\draw[dashed] (0,0) to [out=60, in=120] (1.4,0);
\end{tikzpicture}
=
\begin{tikzpicture}[baseline=(current bounding box.west)]
\fill (-0.6,0) circle[radius=\dotsize]; \draw (-0.6,0) node[below]{$r$};
\fill (0,0) \Square{\dotsize}; \draw (0,0) node[below]{$r_1$};
\fill (0.6,0) \Square{\dotsize}; \draw (0.6,0) node[below]{$r_2$};
\fill (1.2,0) \Square{\dotsize}; \draw (1.2,0) node[below]{$r_3$};
\fill (1.8,0) \Square{\dotsize}; \draw (1.8,0) node[below]{$r_4$};
\draw[dashed] (-0.6,0) to [out=60, in=120] (0.0,0);
\draw[dashed] (0,0) to [out=60, in=120] (0.6,0);
\draw[dashed] (0,0) to [out=60, in=120] (1.2,0);
\draw[dashed] (1.2,0) to [out=60, in=120] (1.8,0);
\end{tikzpicture}~.
\label{eqn:basic_6}
\end{equation}
Since each dashed line carry along a single Weyl potential $\Psi_W$, the number of dashed lines of a diagram is equal to the perturbative order of the corresponding term, e.g. \eqn{eqn:basic_2} is 1st order, \eqn{eqn:basic_3} would be 2nd order and \eqn{eqn:basic_6} would be 4th order etc.

Using these rules, one can systematically construct the deflection to any order in perturbation theory. For the  $\delta x^a{}^{[N]}(r)$ term, draw a set of $N+1$ nodes labeled linearly $r,r_1,r_2,\dots,r_N$. Then the total number of diagrams would be all possible diagrams with $N$ dashed lines linking any two nodes such that all the nodes are linked with the following conditions: (a) there can only be a single line leaving a node to the left, (b) there is at least one line leaving a node to the right and (c) there are no line crossings.  Once all the diagrams are drawn, we then use the rules described above to write down the corresponding intergrals. 

Note that the no-line-crossing condition (c) above is a consequence of the fact that all the integrals are generated by either linked terms or product terms -- which itself is a direct consequence of the fact that integrals are products of deflections which themselves are Green's function solutions of the geodesic equation. If we allow for crossing terms, simple combinatorics tell us that there are $N!$ possible ways of drawing $N$ lines linking $N+1$ nodes with conditions (a) and (b) above. However, imposing condition (c) means that the number of diagrams is less than $N!$.

We will now provide a convenient generating formula for iteratively constructing all the diagrams to all orders in $N$, starting from the two lowest order terms, \eqn{eqn:basic_1} and \eqn{eqn:basic_2}.  Let $\NC{N}$ be the set of the diagrams at $N$-th order, such that $\NC{0}=\{\begin{tikzpicture}
\fill (0,0) \Square{\dotsize}; 
\end{tikzpicture}\}$,
$\NC{1}=\{
\begin{tikzpicture}[baseline=(current bounding box.west)]
\fill (0,0) \Square{\dotsize}; 
\fill (0.6,0) \Square{\dotsize}; 
\draw[dashed] (0,0) to [out=60, in=120] (0.6,0);
\end{tikzpicture}
\}$
,
$\NC{2}=\{
\begin{tikzpicture}[baseline=(current bounding box.west)]
\fill (0,0) \Square{\dotsize}; 
\fill (0.6,0) \Square{\dotsize}; 
\fill (1.2,0) \Square{\dotsize}; 
\draw[dashed] (0,0) to [out=60, in=120] (0.6,0);
\draw[dashed] (0,0) to [out=60, in=120] (1.2,0);
\end{tikzpicture}
,
\begin{tikzpicture}[baseline=(current bounding box.west)]
\fill (0,0) \Square{\dotsize}; 
\fill (0.6,0) \Square{\dotsize}; 
\fill (1.2,0) \Square{\dotsize}; 
\draw[dashed] (0,0) to [out=60, in=120] (0.6,0);
\draw[dashed] (0.6,0) to [out=60, in=120] (1.2,0);
\end{tikzpicture}
\}$ etc. 

Given the collection of known diagrams up to $\NC{N}$, the diagrams of the next order $\NC{N+1}$ are generated by the formula
\begin{equation}
\NC{N+1} = \bigcup_{i=0}^{i=N} \NC{i}\#\NC{N-i}~,\label{eqn:generating_formula}
\end{equation}
where $\bigcup$ denotes ``union of'' and  the product operator $\#$ is defined such that $\NC{i}\#\NC{j}$ denotes linking the left-most node of all the diagrams of $\NC{i}$ with the left-most node of  all the diagrams of $\NC{j}$ with a dashed line.  Note that $\#$ does not commute, i.e. $\NC{i}\#\NC{j}\neq \NC{j}\#\NC{i}$ for $i\neq j$. Since there are no repetitions in the generating formula \eqn{eqn:generating_formula}, if $C_N$ is the number of diagrams in $\NC{N}$, then the total number of diagrams at $N+1$ order is given by the recursion formula
\begin{equation}
C_{N+1} = \sum_{i=0}^{N} C_iC_{N-i}~.\label{eqn:Catalan_derive}
\end{equation}
Here, $C_0=C_1=1$ as $[0]$ and $[1]$ have a single diagram each by construction. Finally, as the recursion formula \eqn{eqn:Catalan_derive} is satisfied by the definition of the \emph{Catalan number}
\begin{equation}
C_{N} = \frac{1}{N+1}\binom{2N}{N}~, \label{eqn:Catalan}
\end{equation}
this proves our assertion in the main text.

\end{document}